\documentclass[pra,twocolumn,showpacs,preprintnumbers,amsmath,amssymb,superscriptaddress,longbibliography]{revtex4-2}
\usepackage{amsmath} 
\usepackage{graphicx}
\usepackage{dcolumn}
\usepackage{bm}
\usepackage{color}
\usepackage{bbm}
\usepackage[colorlinks=true,citecolor=blue,linkcolor=blue,urlcolor=blue]{hyperref}
\usepackage{appendix}
\usepackage{float}
\usepackage{algorithm}
\usepackage{algcompatible}
\usepackage{algpseudocode}
\usepackage{pgffor}

\begin{document}
\title{Extrapolative Quantum Error Mitigation in Continuous-Variable Systems
beyond the Training Horizon}
\author{Jingpeng Zhang}
\thanks{These authors contributed equally to this work.}
\affiliation{School of Physics, Sun Yat-sen University, Guangzhou, Guangdong 510275,
China.}
\author{Shengyong Li}
\thanks{These authors contributed equally to this work.}
\affiliation{Department of Automation, Tsinghua University, Beijing 100084, China.}
\author{Jie Han}
\affiliation{School of Physics, Sun Yat-sen University, Guangzhou, Guangdong 510275,
China.}
\author{Qianchuan Zhao}
\affiliation{Department of Automation, Tsinghua University, Beijing 100084, China.}

\author{Jing Zhang}%
\affiliation{School of Automation Science and Engineering, Xi’an Jiaotong University, Xi’an, 710049, China.}
\author{Ze-Liang Xiang}
\email{zeliangxiang@mail.sysu.edu.cn}

\affiliation{School of Physics, Sun Yat-sen University, Guangzhou, Guangdong 510275,
China.}
\affiliation{State Key Laboratory of Optoelectronic Materials and Technologies, Sun Yat-sen University, Guangzhou 510275, China}
\begin{abstract}
Continuous-variable (CV) quantum systems provide a versatile platform for quantum information processing, in which quantum states can be represented in the quadrature phase space. In realistic implementations, environmental noise, primarily photon loss and dephasing, progressively degrades these states. Machine-learning-based quantum error mitigation (QEM) has recently emerged as a promising approach to suppress such noise; however, existing methods are typically limited to the training horizon and require training data that cover the entire evolution, which is experimentally demanding. Here we introduce a framework for extrapolative quantum error mitigation based on a time-conditioned Swin Transformer. By explicitly embedding the evolution time via adaptive layer normalization, the model learns a correction map that accounts for the continuous accumulation of noise while capturing nonlocal phase-space correlations. Numerical simulations under both Markovian and non-Markovian noise demonstrate accurate state recovery in the long-time regime, where existing approaches deteriorate. Our results establish extrapolative QEM as a practical route to mitigating noise in CV quantum systems without exhaustive training data. 
\end{abstract}
\maketitle

\section{Introduction}
 
In quantum information processing, continuous-variable (CV) architectures provide a practical platform, supporting applications in quantum sensing, optical communication, and quantum computing, among others~\cite{abdoTeleportationEntanglementSwappingContinuousQuantum2025,lloydQuantumComputationContinuousVariables1999,gottesmanEncodingQubitOscillator2001,lvovskyContinuousvariableOpticalQuantumstatetomography2009,adessoContinuousVariableQuantumInformationGaussian2014,andersenHybridDiscreteContinuousvariablequantuminformation2015,yokoyamaUltralargescaleContinuousvariableClusterstatesmultiplexed2013,asavanantGenerationTimedomainmultiplexedTwodimensionalclusterstate2019,grosshansContinuousVariableQuantumCryptographyUsing2002,guQuantumComputingContinuousvariableclusters2009,pirandolaAdvancesQuantumCryptography2020,pirandolaFundamentalLimitsRepeaterlessquantumcommunications2017,takeokaFundamentalRatelossTradeoffopticalquantum2014,holevoEvaluatingCapacitiesBosonicGaussianchannels2001,michaelNewClassQuantumErrorCorrectingCodes2016,leghtasHardwareEfficientAutonomousQuantumMemoryProtection2013,mirrahimiDynamicallyProtectedCatqubitsnewparadigm2014,ofekExtendingLifetimeQuantumbiterror2016a,vasconcelosAllopticalGenerationStatesEncodingqubit2010,andersenContinuousvariableQuantumInformationprocessing2010,arzaniEffectiveDescriptionsBosonicsystemscan2025,braunsteinQuantumInformationContinuousvariables2005,larsenFaultTolerantContinuousVariableMeasurementbasedQuantumComputation2021,maQuantumControlBosonicmodessuperconducting2021}.. In contrast to discrete-variable (DV) systems based on projective measurements with discrete outcomes~\cite{nielsenQuantumComputationQuantum2012}, CV systems encode information in continuous phase-space variables and are naturally measured through Wigner tomography or homodyne detection~\cite{weedbrookGaussianQuantumInformation2012a}. As the states evolve in time, these distributions are progressively reshaped, leading to corresponding changes in their quantum properties. This intrinsic time dependence plays a key role in enabling diverse quantum functionalities 
For example, the preparation of Schr\"{o}dinger cat states in Kerr nonlinear resonators requires sustained evolution under engineered two-photon driving and Kerr interactions~\cite{gravinaCriticalSchrodingerCat2023}.

Environmental noise fundamentally limits the performance of quantum information processing, thereby motivating the development of quantum error mitigation (QEM)~\cite{caiQuantumErrorMitigation2023,giurgica-tironDigitalZeroNoise2020,vandenbergProbabilisticErrorCancellation2023,hugginsVirtualDistillationQuantum2021,loweUnifiedApproachDatadriven2021} strategies. QEM refers to algorithmic techniques that suppress noise-induced bias by classical post-processing of measurement outcomes, enabling the recovery of reliable expectation results without modifying the underlying quantum system. Specifically, in CV platforms, noise arises during dynamical evolution primarily through photon loss and dephasing~\cite{serafiniQuantifyingDecoherenceContinuousvariablesystems2005,wallsQuantumOptics1994,leviantQuantumCapacityCodesbosoniclossdephasing2022,meleQuantumCommunicationBosoniclossdephasingchannel2024}. These noises gradually erode fine structures in phase space and reduce non-classical features that encode useful quantum correlations~\cite{wangDecoherenceDynamicsComplexPhotonStates2009,zurekSubPlanckStructurePhasespaceits2001,ghoshSubPlanckscaleStructuresVibratingmoleculepresence2009,kenfackNegativityWignerFunctionindicatornonclassicality2004}. To counteract this cumulative degradation,  QEM methods are required to retrieve faithful CV states from noisy measurements.

Machine learning has recently emerged as a powerful paradigm for QEM. It learns to suppress errors directly from measurements, achieving reliable information restoration with high accuracy through data-driven approaches. However, existing approaches either focus primarily on discrete-variable systems~\cite{liaoMachineLearningPracticalquantumerror2024,czarnikImprovingEfficiencyLearningbasederrormitigation2025,strikisLearningBasedQuantumErrorMitigation2021} or assume that the training data span the full temporal interval of interest~\cite{liaoNoiseagnosticQuantumErrormitigationdata2025}, commonly referred to as the training horizon. In practice, for CV systems, satisfying this requirement is particularly challenging. Quantum state tomography relies on reconstructing the Wigner function, which requires substantial measurement resources. Moreover, as evolution time increases, noise progressively suppresses fine phase-space structures~\cite{steeleRecoveryQuantumCorrelationsusingmachine2025}, reducing the signal-to-noise ratio and demanding increasingly larger measurement efforts to maintain reconstruction fidelity~\cite{aguiarQuantumLiouvillianTomography2025,olivaQuantumKerrOscillatorsevolutionphase2019,tiunovExperimentalQuantumHomodynetomographymachine2020,varonaLindbladlikeQuantumTomographynonMarkovianquantum2025,whiteNonMarkovianQuantumProcessTomography2022}. These factors render the collection of exhaustive training data across the entire temporal range of interest experimentally costly and often impractical.

Therefore, error mitigation methods capable of learning from experimentally accessible data within a limited training horizon and subsequently extrapolating to mitigate errors at longer, unobserved times would be particularly valuable for CV applications. Achieving such extrapolative capability, however, poses nontrivial challenges arising from intrinsic architectural limitations and how temporal information is incorporated into the network.~\cite{kniggeModellingLongRange2023,ngReassessingLimitationsCNN2021}. Current architectures typically encode temporal information by concatenating time as additional channels to the input Wigner function~\cite{liaoNoiseagnosticQuantumErrormitigationdata2025}, effectively treating a continuous parameter as a discrete index attached to the phase-space representation. While this approach proves adequate for interpolation within the training range, it does not naturally establish an explicit functional dependence on the continuous evolution parameter. Furthermore, as evolution time increases and fine phase-space structures become progressively worn, convolutional operations within CNN-based U-Net architectures~\cite{ronnebergerUNetConvolutionalNetworks2015} are insufficient to recover the faint, long-range correlations necessary to reconstruct these degraded features.

In this work, we develop a neural architecture that explicitly models noise dynamics as a continuous function of evolution time and incorporates nonlocal feature extraction to recover structural correlations from degraded Wigner distributions. Particularly, we adopt the scalable Operator Transformer (scOT) architecture~\cite{herdePoseidonEfficientFoundationModelsPDEs2024}, originally developed for time-dependent partial differential equations (PDEs), and adapt it to open quantum systems. By introducing Adaptive Layer Normalization (AdaLN)~\cite{peeblesScalableDiffusionModelsTransformers2022} as a continuous time-conditioning mechanism, the model captures the dynamical structure of error accumulation, rather than relying on interpolation between discrete time indices. The self-attention mechanism within the U-Net-based framework~\cite{ronnebergerUNetConvolutionalNetworks2015} enables the extractionof weak long-range correlations in phase space, allowing accurate reconstruction even for evolution times well beyond the training regime.

Extensive numerical simulations for both Markovian and non-Markovian noise processes demonstrate that the proposed method maintains high accuracy in regimes where existing approaches deteriorate. These results establish extrapolative quantum error mitigation as a viable approach for recovering faithful quantum states without requiring training data across the full evolution process.


\begin{figure*}[htbp]
\centering %
\noindent\makebox[1\columnwidth]{%
 \includegraphics[width=1.05\textwidth]{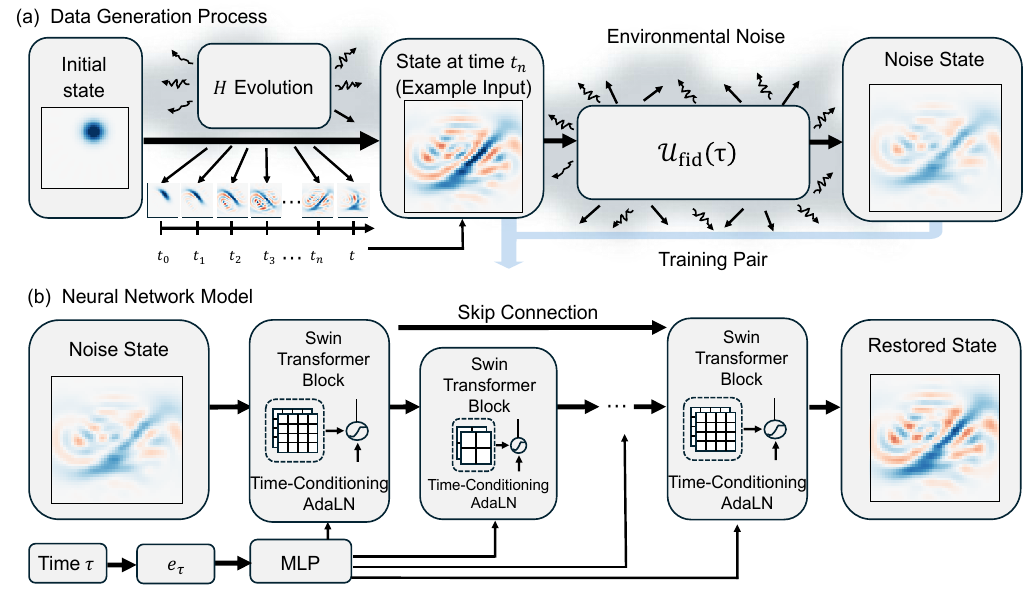}%
} \caption{(a) Training data generation process via fiducial evolution. The system evolves under Hamiltonian $H$ to produce reference states at discrete times $t_{n}$ within the training horizon. Each reference state is then subjected to a fiducial sequence $\mathcal{U}_{\text{fid}}(\tau)$ in the presence of environmental noise, generating the corresponding noisy state. (b) Neural network architecture. Noisy Wigner function inputs are processed by time-conditioned Swin Transformer blocks. The evolution time $\tau$ is embedded through a multi-layer perceptron (MLP) and injected into each Swin transformer block via Adaptive Layer Normalization (AdaLN), enabling the network to model time-dependent noise accumulation.}
\label{fig:workflow} 
\end{figure*}

\section{Methods}
\label{sec:methods}

\subsection{Problem Formulation}
\label{subsec:problem_setup}

We consider the task of learning an extrapolative mapping that reconstructs reference quantum states from noisy Wigner function measurements. Formally, a continuous-variable system evolves under a Hamiltonian $\hat{H}$ in the presence of environmental noise. The state at time $t$, denoted $\rho(t)$, therefore encodes both unitary dynamics and dissipative effects. Additional noise exposure to noise for a duration $\tau$ further transforms the state through a quantum channel $\mathcal{N}_{\tau}$, yielding $\mathcal{N}_{\tau}(\rho(t))$. In phase space, the corresponding noisy observation is represented by the Wigner function $W_{\text{noisy}}(\mathbf{x};\tau)$ with $\mathbf{x}=(q,p)$.

The goal of extrapolative error mitigation is to learn a parametric inverse map:
\begin{equation}
\mathcal{D}_{\theta}:\big(W_{\text{noisy}}(\mathbf{x};\tau),\tau\big)\mapsto W_{\text{target}}(\mathbf{x}),\label{eq:denoising_map}
\end{equation}
where $\mathcal{D}_{\theta}$ denotes a neural network with parameters $\theta$, and $W_{\text{target}}$ represents the reference state prior to the additional noise $\mathcal{N}_{\tau}$. During training, we define a training horizon $T_{\text{train}}$ and restrict the fiducial noise durations in training data to $\tau\in[0,T_{\text{train}}]$. The extrapolation task then requires that $\mathcal{D}_{\theta}$ generalizes reliably to $\tau>T_{\text{train}}$ without explicit training at longer evolution times.

To realize this mapping, we adopt the scOT architecture~\cite{herdePoseidonEfficientFoundationModelsPDEs2024} together with Adaptive Layer Normalization (AdaLN), which introduces an explicit functional dependence on the continuous parameter $\tau$. This design allows the network to learn the temporal structure of error accumulation while using self-attention to capture weak long-range correlations in degraded Wigner distributions. Architectural details are provided in Sec.~\ref{subsec:model}. Training pairs $(W_{\text{noisy}},~W_{\text{target}})$ are generated using the Data Augmentation via Error Mitigation (DAEM) strategy~\cite{liaoNoiseagnosticQuantumErrormitigationdata2025}, detailed in Sec.~\ref{subsec:daem}.


\subsection{Data Generation via Fiducial Processes}
\label{subsec:daem}

The mapping defined in Eq.~\eqref{eq:denoising_map} requires paired training data $(W_{\text{noisy}},W_{\text{target}})$. A straightforward approach would require access to ideal noiseless states as targets, which is experimentally infeasible for systems undergoing complex dynamics. To circumvent this limitation, we adopt the Data Augmentation via Error Mitigation (DAEM) strategy introduced by Liao \textit{et al.}~\cite{liaoNoiseagnosticQuantumErrormitigationdata2025}, which generates training pairs through controlled fiducial operations without requiring perfectly isolated quantum states.

The DAEM procedure separates controllable Hamiltonian dynamics from environmental noise. Consider a system that has evolved under the Hamiltonian $\hat{H}$ in the presence of environmental noise up to time $t_{k}\in[0,T_{\text{train}}]$, producing the state $\rho(t_{k})$.
This state, which we refer to as the pre-fiducial state, serves as the training target. Importantly, $\rho(t_{k})$ is not an ideal pure state; rather, it already contains the environmental noise accumulated during the evolution to time $t_{k}$, representing the experimentally accessible approximation to the system's dynamical state.

To generate the corresponding noisy input, we apply a fiducial sequence of total duration $\tau$ to $\rho(t_{k})$. This sequence consists of forward evolution under $+\hat{H}$ for a duration $\tau/2$, followed by backward evolution under $-\hat{H}$ for the same duration, 
\begin{equation}
\mathcal{U}_{\text{fid}}(\tau)=e^{+i\hat{H}\tau/2}\,e^{-i\hat{H}\tau/2}.
\end{equation}
In the absence of environmental noise, these operations cancel exactly ($\mathcal{U}_{\text{fid}}=\mathbb{I}$), leaving the state $\rho(t_{k})$ unchanged. Under realistic conditions, however, noise acts continuously during the fiducial sequence, leading to additional decoherence described by a quantum channel $\mathcal{N}_{\tau}$: 
\begin{equation}
\rho_{\text{noisy}}^{(\tau)}(t_{k})=\mathcal{N}_{\tau}\big[\rho(t_{k})\big].
\end{equation}

This construction produces training pairs in which the input is the Wigner function of $\rho_{\text{noisy}}^{(\tau)}(t_{k})$, carrying the additional fiducial noise $\tau$, while the target is the Wigner function of $\rho(t_{k})$, i.e., the pre-fiducial state. By varying both the evolution time $t_{k}$ (which determines the reference state) and the fiducial duration $\tau$ (which controls the noise dynamics), we obtain the dataset 
{\footnotesize
\begin{equation}
\mathcal{D}=\left\{ \big(W_{\text{noisy}}^{(\tau)}(t_{k}),\,\tau,\,W_{\text{ref}}(t_{k})\big)\,\middle|\,t_{k}\in[0,T_{\text{train}}],\,\tau\in[0,T_{\text{train}}]\right\},
\end{equation}
}
where $W_{\text{ref}}(t_{k})$ denotes the Wigner function of the noisy (non-ideal) reference state at time $t_{k}$.

A key advantage of this construction is that it does not require access to ideal noiseless states. Instead, the model learns to remove the excess noise introduced by the fiducial process, recovering the pre-fiducial state $\rho(t_{k})$ rather than reconstructing an unattainable ideal state. This makes the framework experimentally feasible even when exact quantum control or noiseless simulations are unavailable.

Algorithm~\ref{alg:data_gen} summarizes the data-generation procedure. While the DAEM protocol originates from prior work
on CNN-based error mitigation~\cite{liaoNoiseagnosticQuantumErrormitigationdata2025}, its integaration with time-conditioned architectures enables the extrapolative capabilities central to the present work. 
\begin{algorithm}[H]
\caption{Training Data Generation via DAEM}
\label{alg:data_gen}
\begin{algorithmic}
\Require Hamiltonian $\hat{H}$,
initial state $\rho_{0}$, training horizon $T_{\text{train}}$, evolution
time samples $\{t_{k}\}$, noise durations $\{\tau_{j}\}$
\Ensure Training dataset $\mathcal{D}$
\State Initialize $\mathcal{D}\leftarrow\emptyset$
\For{each evolution time $t_{k}\in[0,T_{\text{train}}]$}
\State Evolve system to $t_{k}$ under $\hat{H}$ and noise: $\rho(t_{k})=\mathcal{E}_{t_{k}}(\rho_{0})$
\State Measure reference Wigner function: $W_k=\mathcal{W}[\rho(t_{k})]$
\For{each noise duration $\tau_{j}\in[0,T_{\text{train}}]$}
\State Apply fiducial sequence: \State$\rho_{\text{noisy}}=\mathcal{U}_{\text{fid}}(\tau_{j})\rho(t_{k})\,\mathcal{U}_{\text{fid}}^{\dagger}(\tau_{j})$
\State Measure noisy Wigner function: $W_{k}^{(j)}=\mathcal{W}[\rho_{\text{noisy}}]$
\State Append $W_{k}^{(j)},\tau_{j},W_{k}$ to $\mathcal{D}$
\EndFor
\EndFor
\State \Return $\mathcal{D}$
\end{algorithmic}
\end{algorithm}


\subsection{Time-Conditioned Neural Architecture}
\label{subsec:model}

To realize the mapping in Eq.~\eqref{eq:denoising_map}, we adopt the scOT architecture~\cite{herdePoseidonEfficientFoundationModelsPDEs2024}, originally developed for time-dependent PDEs, and adapt its time-conditioning mechanisms for dissipative quantum dynamics.

The model follows a hierarchical Swin Transformer U-Net architecture that processes Wigner functions through an encoder-decoder pipeline with skip connections, enabling multi-scale feature extraction while preserving fine phase-space structure. Although both CNN- and Transformer-based U-Nets provide global receptive fields, the self-attention mechanism offers enhanced capability to capture weak long-range correlations in degraded phase-space distributions. As dissipation progresses, intricate features in the Wigner function become progressively attenuated and survive primarily as subtle nonlocal patterns. The dynamic attention mechanism enables the network to identify and amplify these residual correlations even in low-contrast distributions.

Temporal extrapolation is achieved using Adaptive Layer Normalization (AdaLN)~\cite{peeblesScalableDiffusionModelsTransformers2022,songScoreBasedGenerativeModelingStochasticDifferential2020}, which treats the duration $\tau$ as a continuous conditioning variable. The scalar $\tau$ is embedded into a multiscale feature vector $\mathbf{e}_{\tau} \in \mathbb{R}^{d}$ through a nonlinear embedding network. 
Within each Swin Transformer block, the normalization layer is modulated as
\begin{equation}
\text{AdaLN}(\mathbf{x},\mathbf{e}_{\tau})=\gamma(\mathbf{e}_{\tau})\odot\frac{\mathbf{x}-\mu(\mathbf{x})}{\sigma(\mathbf{x})}+\beta(\mathbf{e}_{\tau}),
\end{equation}
where $\gamma$ and $\beta$  are scale and shift parameters generated by a multilayer perceptron (MLP) acting on the time embedding. This conditioning introduces an explicit functional dependence of the network on the continuous evolution parameter $\tau$, allowing the model to learn a smooth family of operators $\mathcal{D}_{\theta}(\cdot;\tau)$ that generalize beyond the training horizon. Consequently, the network can infer how the correction operation varies continuously with noise duration, enabling extrapolative error mitigation beyond the interpolation regime. 
the interpolation regime. 



\subsection{Training and Inference}
\label{subsec:training}

The network parameters $\theta$ are optimized by minimizing the $l_1$ distance between the predicted and reference Wigner functions, 
\begin{equation}
\mathcal{L}(\theta)=\mathbb{E}_{(W_{\text{noisy}}^{(\tau)},\tau,W_{\text{ref}})\sim\mathcal{D}}\left[\big\|\mathcal{D}_{\theta}(W_{\text{noisy}}^{(\tau)},\tau)-W_{\text{ref}}\big\|_{1}\right],
\end{equation}
where the expectation is taken over the training dataset generated using the DAEM protocol. During training, both the evolution time $t_{k}$ and the noise duration $\tau$ are restricted to the training horizon $[0,T_{\text{train}}]$. Therefore, the model never observes corruption beyond this temporal window, and generalization to longer times must arise from the learned continuous dependence on $\tau$ rather than memorization of long-time examples.

It is important to distinguish between the training target and the ultimate inference objective. As detailed in Sec.~\ref{subsec:daem}, the training target $W_{\text{ref}}$ corresponds to the pre-fiducial state $\rho(t_k)$, which already contains environment noise accumulated during forward evolution to time $t_{k}$, but excludes the additional fiducial noise $\mathcal{N}_{\tau}$ introduced by the DAEM sequence. The network therefore learns to remove the additional noise associated with the fiducial process, recovering the noisy baseline rather than reconstructing an ideal noiseless state.

During inference, however, the trained model is used to perform full quantum error mitigation. For a given test time $t_{\text{test}}$ (including the extrapolation regime $t_{\text{test}}>T_{\text{train}}$), the model receives a noisy Wigner function generated by pure forward evolution under environmental noise and the corresponding time parameter $\tau_{\text{test}}$($\tau_{\text{test}}=t_{\text{test}}$during forward evolution). The network applies the denoising operation $\mathcal{D}_{\theta}(\cdot;\tau_{\text{test}})$ through the AdaLN mechanism. In this regime, the target shifts to the ideal noiseless state that would result from Hamiltonian evolution alone in the absence of environmental noise. It represents a significant generalization beyond the training task: the model must extrapolate from removing fiducial noise (DAEM) to mitigating the full environmental noise, while operating at evolution times beyond the training horizon~\cite{songTimelyGPTExtrapolatableTransformer2023,sunLengthExtrapolatableTransformer2023}.
For $\tau_{\text{test}}\leq T_{\text{train}}$, the model interpolates within the training regime; for $\tau_{\text{test}}>T_{\text{train}}$, it extrapolates beyond the observed horizon by following the learned noise trajectory.

The performance of error mitigation is quantified by the cosine similarity between the reconstructed Wigner function and the reference state~\cite{manningIntroductionInformationRetrieval2008,sellierSensitivityStudyWignerMonteCarlo2015}:
\begin{equation}
S=\frac{\langle\hat{W}_{\text{pred}},W_{\text{target}}\rangle}{\|\hat{W}_{\text{pred}}\|\cdot\|W_{\text{target}}\|},
\end{equation}
where $\langle\cdot,\cdot\rangle$ denotes the inner product over the phase-space grid and $\hat{W}_{\text{pred}}$ represents the output of the neural network. This metric measures the structural fidelity of the reconstructed quantum state, with values approaching unity indicating successful recovery of the target state. Detailed hyperparameters and optimization settings are provided in the Supplementary Material. 


\section{Numerical Demonstrations}
\label{sec:results} 

\subsection{Overview and Experimental Protocols}
\label{subsec:setup}

We evaluate the extrapolative error mitigation performance of the proposed time-conditioned Swin Transformer against the CNN U-Net baseline~\cite{liaoNoiseagnosticQuantumErrormitigationdata2025} under two distinct noise regimes: Markovian and non-Markovian. In both cases, photon loss and dephasing constitute the dominant environmental noise processes. The Markovian setting treats these mechanisms as memoryless channels, whereas the non-Markovian case incorporates history-dependent effects through the reaction-coordinate (RC) formalism~\cite{nazirReactionCoordinateMapping2018}. 

The experiments are designed to evaluate the extrapolative error-mitigation capability of the proposed architecture beyond the training horizon $T_{\text{train}}$ across diverse Hamiltonians and noise regimes. To ensure a fair comparison, both models adopt the same encoder-decoder U-Net backbones with skip connections and are trained on identical datasets with the same training horizon $T_{\text{train}}$. For each experiment, multiple initial coherent states are evolved under a common Hamiltonian $\hat{H}$ with a forward loss rate $\kappa_{\text{forward}}$. Each state is then subjected to a range of loss rates $\kappa_i \in [\kappa_{\min}, \kappa_{\max}]$ during the DAEM sequence, producing diverse training pairs.


\subsubsection{Markovian Regime: Multi-channel Snapshot Protocol}
\label{subsubsec:markov_protocol}

For Markovian dynamics, we adopt a multichannel snapshot protocol consistent with the CNN U-Net baseline~\cite{liaoNoiseagnosticQuantumErrormitigationdata2025}. The input consists of five channels, each representing the Wigner function of the same reference state (evolved from a single initial coherent state with $\kappa_\text{forward}$) at a different loss rate $\kappa_i$ within the DAEM sequence. In the baseline model, these five noise channels are concatenated with a sixth channel encoding the scalar evolution time $\tau$, spatially replicated to match the Wigner-function grid. In contrast, our approach embeds $\tau$ through AdaLN, treating it as a continuous conditioning variable.

This multichannel design provides simultaneous observations of the state under different noise intensities, facilitating disentanglement of systematic noise effects from random fluctuations during training. We emphasize that both architectures receive identical physical information regarding the quantum state under varying noise conditions.

The training target is the pre-fiducial state $\rho_{\text{ref}}(t)$ obtained from forward evolution under the Hamiltonian and environmental noise up to time $t$. Thus, the training task is to remove additional noise introduced by the high-loss channels and recover the baseline noisy state.

During inference, we employ a direct reconstruction strategy. For a test time $t_{\text{test}}$, five-channel input are generated by evolving the system under five distinct loss rates $\kappa_\text{forward} \in [\kappa_{\min}, \kappa_{\max}]$, consistent with the DAEM training protocol. The corresponding time parameter $\tau_{\text{test}}$ is embedded via AdaLN, and the network performs the denoising operation in a single forward pass, directly outputting the corrected Wigner function without iterative refinement.


\subsubsection{Non-Markovian Regime: Iterative Step-wise Protocol}
\label{subsubsec:nonmarkov_protocol}

Non-Markovian dynamics exhibits path-dependent behavior arising from memory effects in the system-environment interaction~\cite{devegaDynamicsNonMarkovianOpenquantumsystems2017,breuerMeasureDegreeNonMarkovianBehaviorQuantum2009,rivasEntanglementNonMarkovianityQuantumEvolutions2010,huQuantumBrownianMotiongeneralenvironment1992,richterPhasespaceMeasuresInformationflowopen2024}. 
Unlike Markovian dynamics, where the Wigner function typically undergoes relatively uniform broadening, memory effects induce nontrivial deformations in the phase-space distribution that accumulate over time. Accurately modeling such dynamics, therefore, requires an explicit representation of the temporal evolution that preserves the history-dependent structure of the state trajectory.

To address this challenge, we adopt an iterative stepwise reconstruction protocol that models the evolution through sequential temporal increments. During training, we construct sample pairs consisting of states evolved under identical conditions but at different time points $\tau_1$ and $\tau_2$ ($\tau_1 > \tau_2$). The network input is a two-channel tensor comprising the Wigner function $W_{\text{noisy}}(\mathbf{x};\tau_1)$ and a temporal channel encoding the step duration $\Delta\tau$ ($\Delta\tau=\tau_1-\tau_2$), represented as a spatial uniform map matching the Wigner grid. 
Through AdaLN conditioning, the denoising transformation depends on both the current time $\tau_1$ and the step size $\Delta\tau$, allowing the network to learn corrections that reflect the system's dynamical history.

This formulation enables the network to approximate an infinitesimal denoising operator $\mathcal{D}_{\theta}(\cdot;\tau,\Delta\tau)$ that maps the noisy state at $\tau_1$ toward the reference state at $\tau_2$. Training on such pairs helps disentangle cumulative noise effects from the underlying Hamiltonian evolution.

During inference, the state trajectory is reconstructed iteratively. Given a measured state with total noise exposure $\tau_{\text{total}}$, the model is applied sequentially using small increments $\Delta\tau$ to approximate the continuous trajectory. At each step, the current Wigner estimate, the updated time $\tau_{\text{total}} - n\Delta\tau$ ( with $n$ denoting the iteration index), and the step size $\Delta\tau$ are provided as inputs. The AdaLN layers adapt to both the accumulated evolution time and the current increment, enabling the model to approximate history-dependent dynamics through a sequence of discrete updates.

For comparison, the CNN U-Net baseline follows the configuration of Ref.~\cite{liaoNoiseagnosticQuantumErrormitigationdata2025}. It employs the same six-channel snapshot architecture used in the Markovian experiments (five Wigner channels plus one time channel) for both training and inference.


\subsubsection{Model Configurations and Generalization Protocol}
\label{subsubsec:config}

To ensure that the model learns underlying physical laws rather than memorizing specific trajectories, we adopt a strict generalization protocol over initial states. The training dataset consists of evolution trajectories generated from 20 randomly sampled coherent states within the phase-space region $|\alpha|\leq 2$, each evolved under the corresponding Hamiltonian. For each initial state, we first evolve the system under the Hamiltonian with a fixed forward loss rate $\kappa_{\text{forward}}$ to produce the reference trajectory. The DAEM procedure is then applied using five distinct loss rates $\kappa_i$ ($i=1,\dots,5$) spanning the interval $[\kappa_{\min},\kappa_{\max}]$, generating noisy inputs corresponding to different noise intensities. This procedure yields a total of 100 training trajectories (20 initial states $\times$ 5 loss rates). 

During testing, we test the model on novel coherent states randomly drawn from the same phase-space region but excluded from the training set, thereby verifying generalization to previously unseen quantum states.

The training horizon is defined as $T_{\text{train}}=1.0$ (arbitrary units), and the model is trained exclusively on data with evolution times $t\in[0,T_{\text{train}}]$. The test horizon extends to $t\in[0,2T_{\text{train}}]$, encompassing both the interpolation regime $[0,T_{\text{train}}]$, where the model validates its ability to reproduce training distributions, and the extrapolation regime $(T_{\text{train}},2T_{\text{train}}]$, where it encounters evolution times and corresponding phase-space distributions not observed during training. All Wigner functions are discretized on a $48\times48$ grid over the phase-space domain $[-4,4]\times[-4,4]$.

As detailed in Sec.~\ref{subsubsec:markov_protocol} and Sec.~\ref{subsubsec:nonmarkov_protocol}, the input-channel configuration differs between noise regimes. For Markovian dynamics, the network receives five channels corresponding to Wigner functions generated with different loss rates. For non-Markovian dynamics, the input consists of two channels: the Wigner function and a temporal coordinate $\Delta\tau$. In both cases, the noise duration $\tau$ and the step size $\Delta\tau$ are incorporated through adaptive layer normalization, allowing the network to modulate its internal transformations according to the continuous noise-exposure parameters. This design enables the model to learn denoising operations that vary smoothly with evolution time. Key hyperparameters are summarized in Table~\ref{tab:hyperparams}.

\begin{table}[h]
\centering \caption{Key hyperparameters and input configurations of the time-conditioned
Swin Transformer.}
\label{tab:hyperparams} %
\begin{tabular}{lcc}
\hline 
Parameter  & Markovian  & Non-Markovian \tabularnewline
\hline 
Embedding dimension  & 48  & 48 \tabularnewline
Attention window size  & $8\times8$  & $6\times6$ \tabularnewline
Batch size  & 1024  & 1024 \tabularnewline
Training horizon $T_{\text{train}}$  & 1.0  & 1.0 \tabularnewline
Test horizon  & $2T_{\text{train}}$  & $2T_{\text{train}}$ \tabularnewline
Input channels  & 5  & 2 (Wigner + $\Delta\tau$) \tabularnewline
Inference strategy  & Direct  & Iterative ($\Delta\tau$ steps) \tabularnewline
\hline 
\end{tabular}
\end{table}




\subsection{Markovian Dynamics: Amplitude Calibration in Long-time Evolution}
\label{sec:markov}

We evaluate extrapolative mitigation performance using two representative Hamiltonians under environmental noise comprising photon loss ($\propto\hat{a}$) and dephasing ($\propto\hat{a}^{\dagger}\hat{a}$), with the dephasing strength set to one-twentieth of the loss rate. The system dynamics are governed by the Lindblad master equation:
\begin{equation}
\frac{d\rho}{dt} = -i[\hat{H}, \rho] + \mathcal{D}[\sqrt{\kappa}\hat{a}]\rho + \mathcal{D}[\sqrt{\kappa/20}\,\hat{a}^{\dagger}\hat{a}]\rho,
\label{eq:lindblad_markov}
\end{equation}
where $\mathcal{D}[\hat{L}]\rho = \hat{L}\rho\hat{L}^{\dagger} - \frac{1}{2}\{\hat{L}^{\dagger}\hat{L}, \rho\}$ denotes the Lindblad dissipator and $\kappa$ is the photon loss rate. 

We consider two Hamiltonians representing different dynamical regimes. The Kerr Hamiltonian, $\hat{H}=1.2\hat{a}^{\dagger 2}\hat{a}^{2}$, provides a standard benchmark for simple nonlinear dynamics. The driven squeezing Hamiltonian, 
\begin{equation}
\hat{H}=-\Delta\hat{a}^{\dagger}\hat{a}+K\hat{a}^{\dagger 2}\hat{a}^{2}-P_{0}(\hat{a}+\hat{a}^{\dagger})
\end{equation} 

with $K=0.5$, $\Delta=1.0$, and $P_{0}=1.0$, generates more complex trajectories involving coherent displacements and quadrature squeezing. Such dynamics are widely studied in quantum metrology and precision measurement protocols~\cite{chavez-carlosQuantumSensingKerrparametricoscillators2024,caiQuantumSqueezingAmplificationweakKerr2025,guoQuantumMetrologySqueezedKerroscillator2024}, where state sensitivity to loss necessitates effective error mitigation. 

Both training and testing employ five-channel inputs corresponding to discrete loss rates $\kappa\in\{0.3,0.4,0.5,0.6,0.7\}$, with the forward evolution rate set to $\kappa_{\text{forward}}=0.3$ for the Kerr case. For the squeezing Hamiltonian, the loss rates are rescaled by $1/3$, yielding $\kappa\in\{0.1,0.133,0.167,0.2,0.233\}$ with $\kappa_{\text{forward}}=0.1$). During testing, the five-channel inputs span the same loss-rate ranges as in training ($\kappa\in\{0.3,0.4,0.5,0.6,0.7\}$ for Kerr or $\{0.1,0.133,0.167,0.2,0.233\}$ for squeezing).

\begin{figure}[htbp]
\centering \includegraphics[width=1\columnwidth,height=0.68\columnwidth]{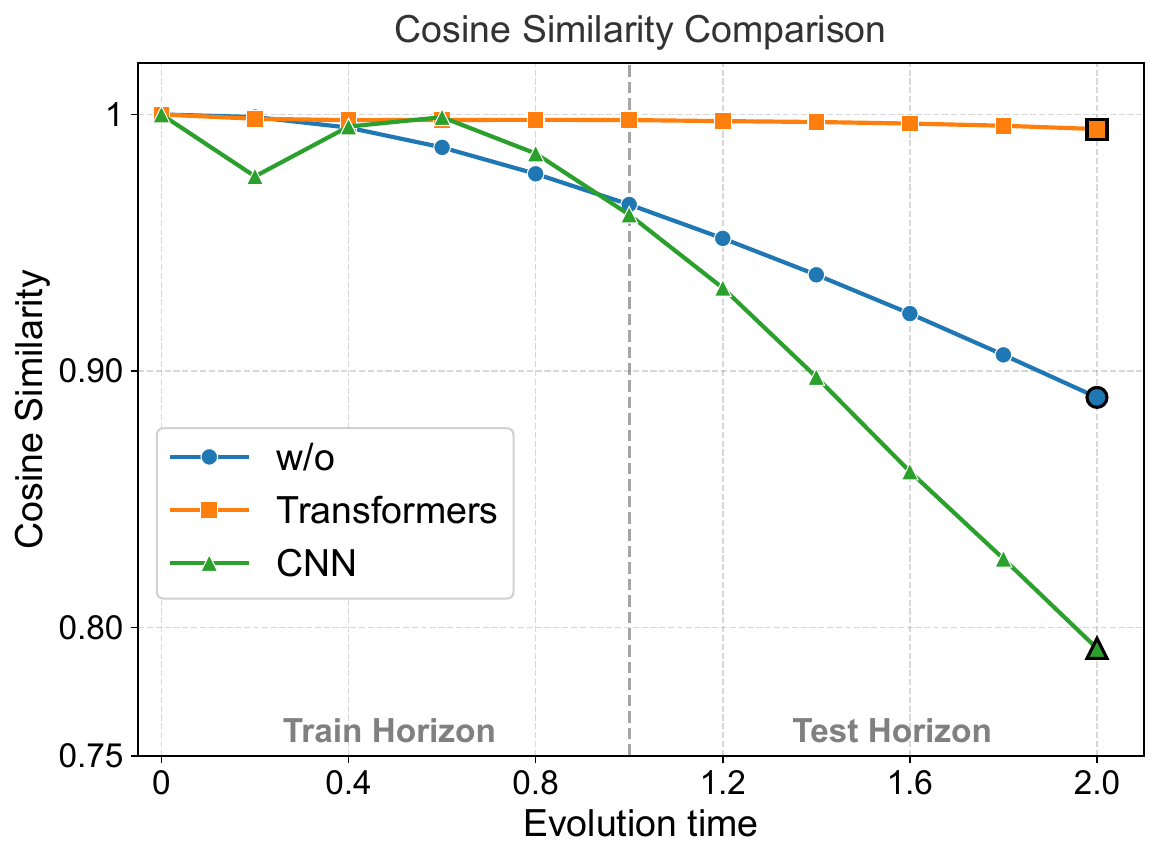}
\noindent\makebox[1\columnwidth]{%
 \includegraphics[width=1.05\columnwidth,height=0.32\columnwidth]{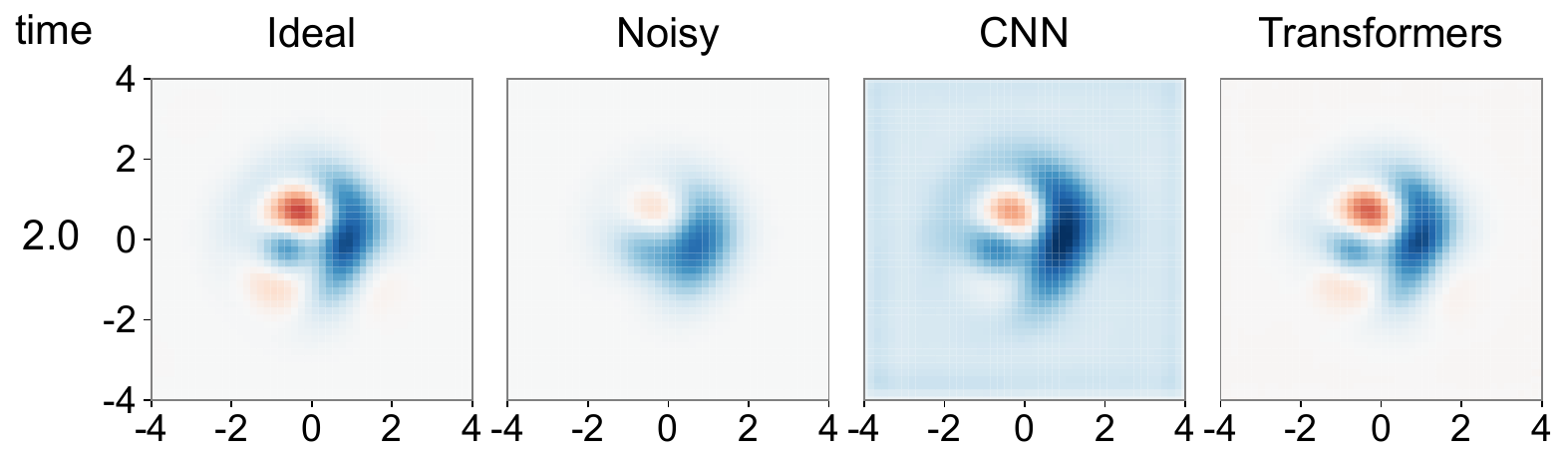}%
} \caption{Cosine similarity for Kerr nonlinearity ($K=1.2$). Training uses loss rates $\kappa\in\{0.3,0.4,0.5,0.6,0.7\}$ with test loss rate $\kappa\in\{0.3,0.4,0.5,0.6,0.7\}$. Within the training horizon $T_{\text{train}}=1.0$, both models achieve high similarity. Beyond this range, the CNN U-Net (green) degrades rapidly due to amplitude miscalibration, showing numerical overflow and spurious background excitation (red arrows) at $t=2.0$. In contrast, the time-conditioned Swin Transformer (orange) maintains stable reconstruction by dynamically adapting the normalization scale through AdaLN.}
\label{fig:kerr} 
\end{figure}

For the Kerr dynamics, the test initial state is a coherent state $\vert\alpha\rangle$ with $\alpha=0.80-0.45i$ (randomly chosen). Within the training horizon ($t\leq1.0$), both architectures achieve high similarity ($>0.98$). In the extrapolation regime ($t>1.0$), however, the CNN U-Net exhibits rapid degradation, dropping to $\sim$0.79 at $t=2.0$, whereas the Swin Transformer maintains a similarity of about $0.99$ (Fig.~\ref{fig:kerr}). This failure of the CNN model arises from amplitude mismatch. As photon loss accumulates over time, the Wigner function is progressively washed out, with peaks smoothed away. The CNN's static encoding cannot adapt to feature scaling during temporal evolution, leading to numerical overflow and spurious background excitation when $W\approx0$ is expected. In contrast, the Swin Transformer employs AdaLN to dynamically adjust normalization parameters based on the evolution time $\tau$, allowing the denoising strength to scale appropriately with the degree of corruption. 

\begin{figure}[h]
\centering \includegraphics[width=1\columnwidth,height=0.68\columnwidth]{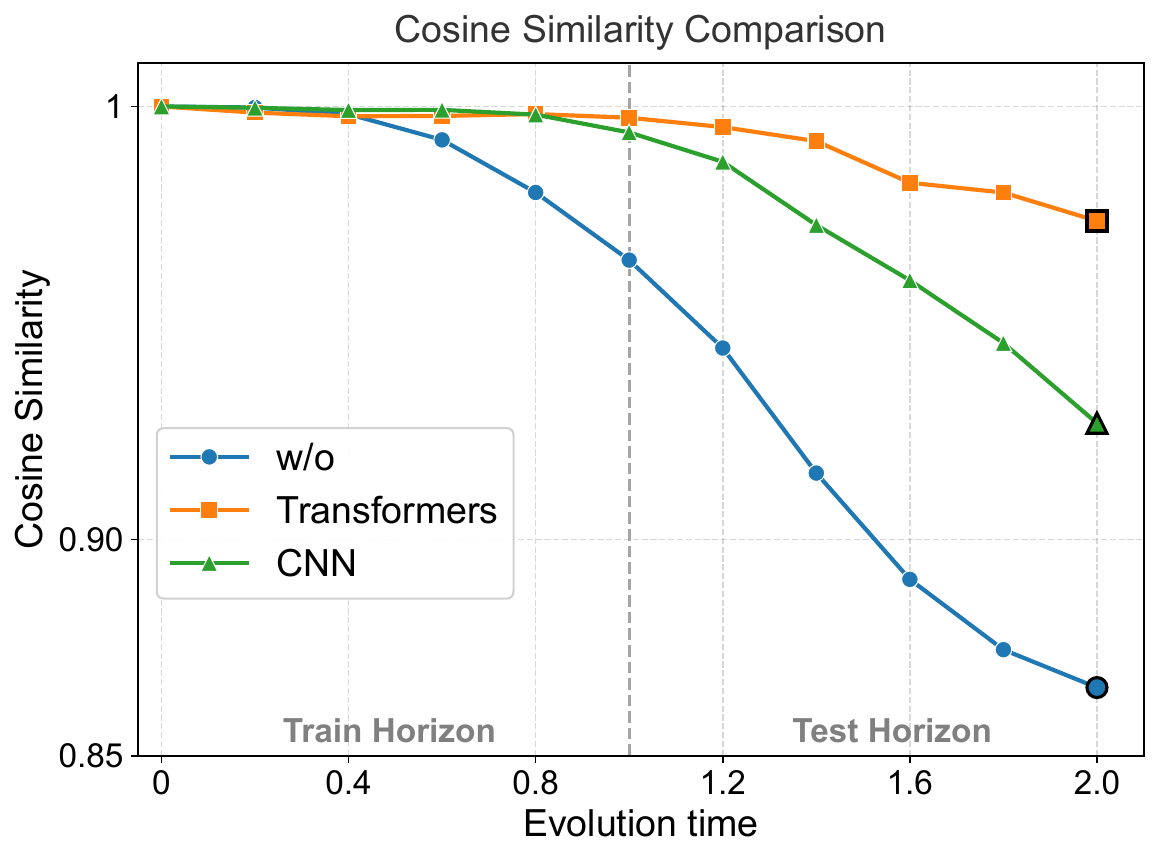}
\noindent\makebox[1\columnwidth]{%
\includegraphics[width=1.05\columnwidth,height=0.32\columnwidth]{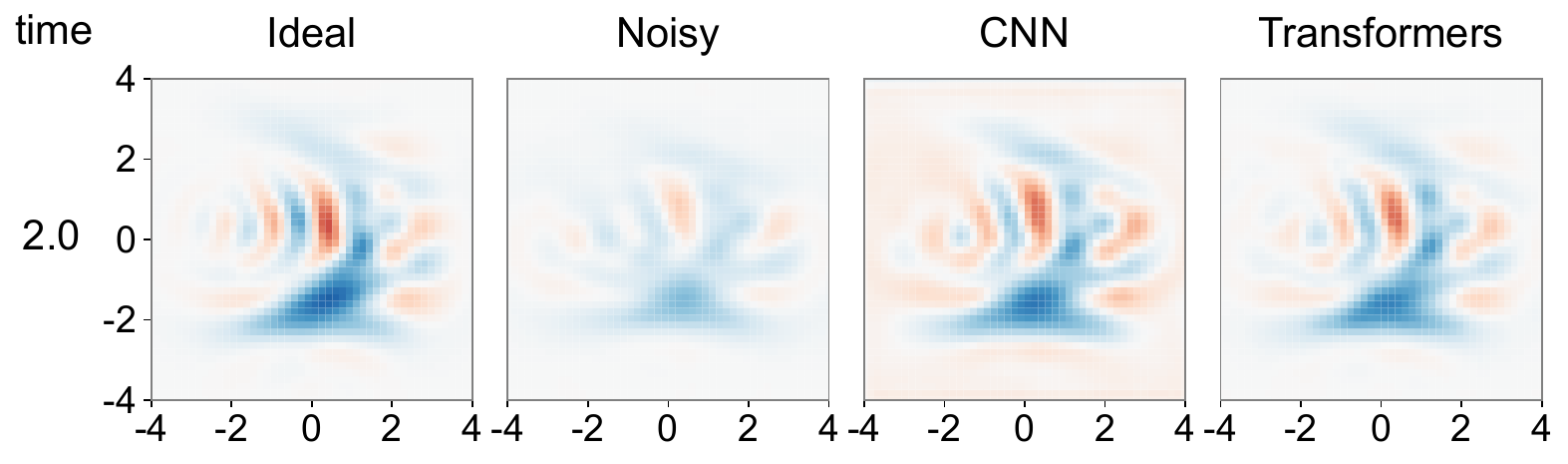}%
} \caption{Extrapolation for driven squeezing dynamics (training loss rates rescaled by $1/3$, testing at $\kappa\in\{0.1,0.133,0.167,0.2,0.233\}$). The reduced dissipation mitigates amplitude errors; the CNN U-Net (green) achieves higher similarity ($\sim$0.92) than in Kerr, because cosine similarity is insensitive to global amplitude deviations when geometric structure is preserved. The Swin Transformer (orange) maintains superior similarity ($\sim$0.97).}
\label{fig:squeezing} 
\end{figure}

For the driven squeezing dynamics, the test initial state is $\vert\alpha\rangle$ with $\alpha=0.63+1.23i$ (randomly chosen). The smaller loss rate ($\kappa_{0}=0.1$) reduces the severity of amplitude errors. Nevertheless, the CNN baseline still exhibits background excitation in outer phase-space regions. Interestingly, the quantitative gap between the models becomes smaller ($\sim$0.92 vs $\sim$0.97 at $t=2.0$, Fig.~\ref{fig:squeezing}), despite the continued presence of amplitude miscalibration. This apparent CNN improvement reflects the cosine metric's insensitivity to global amplitude scaling when geometric structure (squeezing direction, ellipticity) is preserved. Because squeezing primarily produces regular geometric distortions that CNN convolutional biases capture effectively, masking amplitude deviations in the scalar score. Nevertheless, the Swin Transformer provides more reliable physical reconstruction through explicit temporal conditioning.

\subsection{Non-Markovian Dynamics: Memory Effects and Fine Feature Preservation}
\label{sec:nonmarkov}

We extend our analysis to non-Markovian environments, where environmental memory introduces history-dependent noise dynamics. The dominant noise channels remain photon loss ($\propto\hat{a}$) and dephasing ($\propto\hat{a}^{\dagger}\hat{a}$, with one-twentieth the strength of loss), now mediated through the reaction-coordinate model with memory correlation time $\tau_{c}\sim\kappa^{-1}$.The system Hamiltonian $H_{sys}$ is the driven squeezing Hamiltonian. As in the Markovian experiments, the training loss rates are $\kappa\in\{0.3,0.4,0.5,0.6,0.7\}$ with $\kappa_{forward}=0.3$. Extrapolation is evaluated at a test rate of $\kappa_{\text{forward}}=0.3$, where memory effects induce pronounced path-dependent deformations of the phase-space distribution. 

For this regime, we employ the iterative stepwise reconstruction protocol described in Sec.~\ref{subsubsec:nonmarkov_protocol}, which reconstructs the trajectory by sequentially applying corrections conditioned on the accumulated time $\tau$. The CNN U-Net still receives the standard six-channel input (five loss rates $\kappa_\text{forward}\in\{0.3,0.4,0.5,0.6,0.7\}$ plus time), whereas our method employs the iterative protocol with access to only the loss rate $\kappa_\text{forward}=0.3$ and the step size $\Delta\tau$.

\begin{figure}[t]
\centering \includegraphics[width=1\columnwidth,height=0.68\columnwidth]{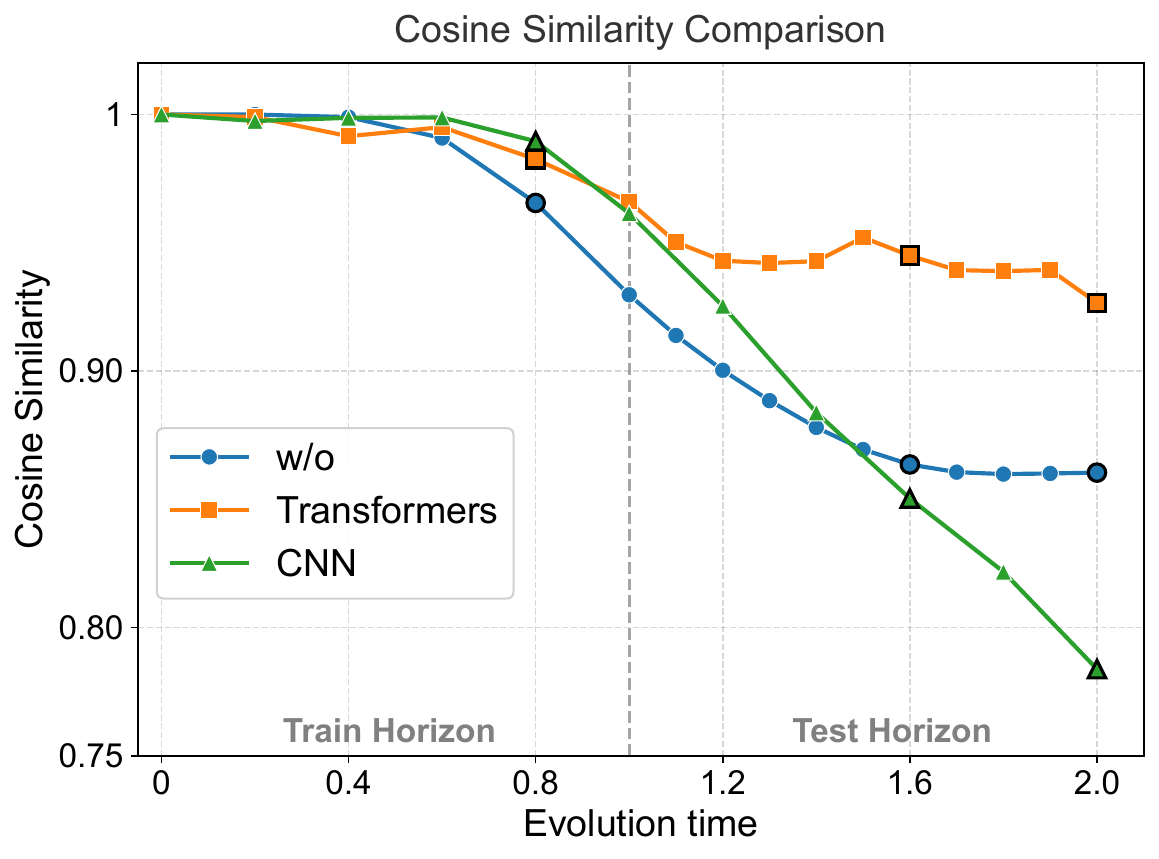}
\noindent\makebox[1\columnwidth]{%
 \includegraphics[width=1.13\columnwidth]{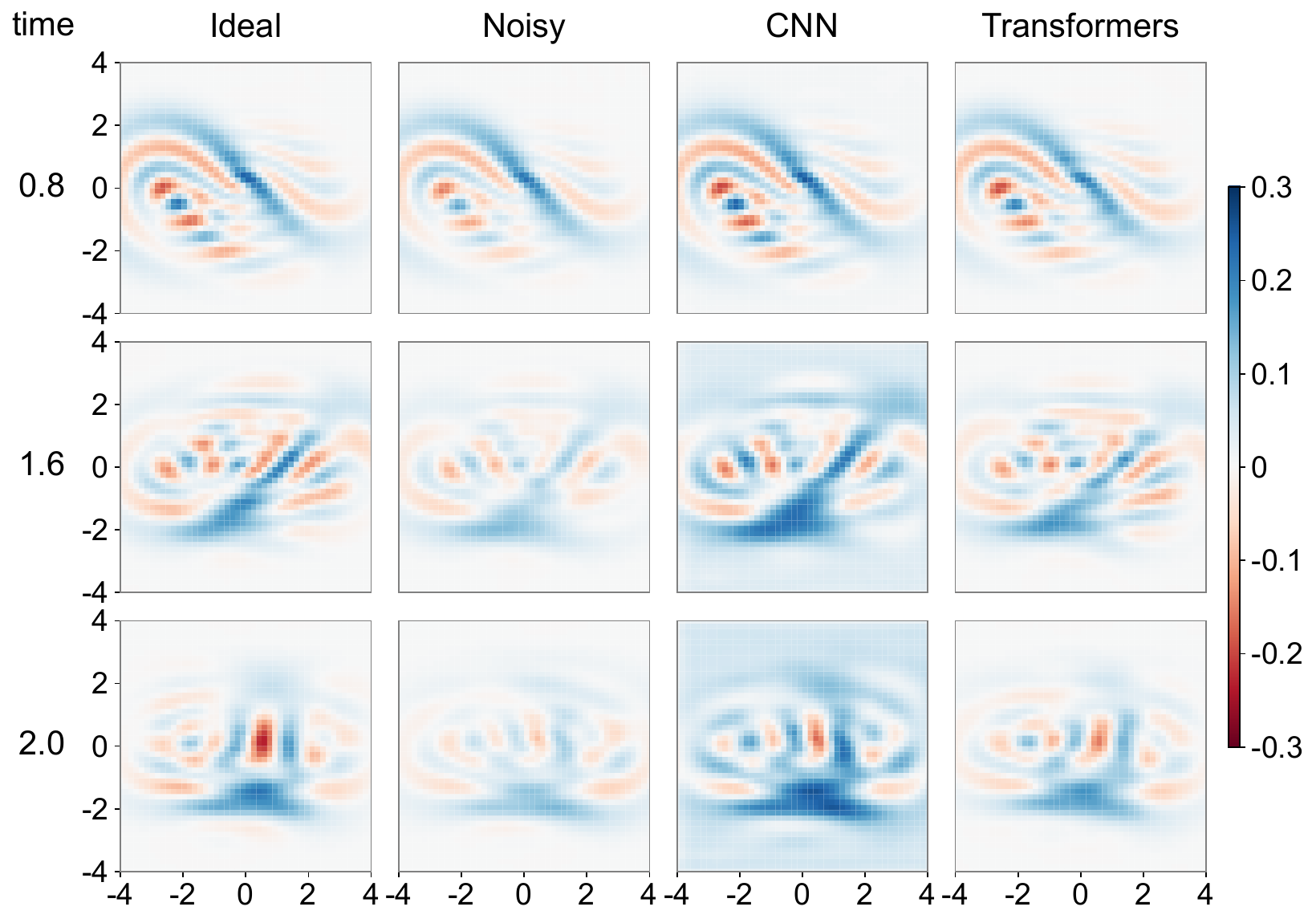}%
} \caption{Performance under non-Markovian dynamics (training loss rate at $\kappa\in\{0.3,0.4,0.5,0.6,0.7\}$, testing at $\kappa_{\text{forward}}=0.3$). The dashed line marks the training horizon $t=1.0$. Beyond this range ($t>1.0$), the CNN U-Net (green triangles) degrades due to  shape distortion and loss of fine  structure details. In contrast, the Swin Transformer (orange squares) better preserves structural correlations through sensitive extraction, maintaining $\sim$0.92 similarity with characteristic non-monotonic behavior.}
\label{fig:nm_squeezing} 
\end{figure}

The test initial state is $\vert\alpha\rangle$ with $\alpha=0.34+0.97i$ (chosen randomly). Within the interpolation regime ($t\leq1.0$), both architectures maintain high similarity (similarity $>0.95$). Beyond the training horizon ($t>1.0$), however, their behaviors diverge significantly (Fig.~\ref{fig:nm_squeezing}). The CNN U-Net exhibits monotonic degradation to $\sim$0.78 at $t=2.0$, showing not only amplitude overflow (spurious background excitation) but also progressive shape distortion, with fine phase-space details being obscured. In contrast, the Swin Transformer maintains similarity around $0.93$, with non-monotonic performance variations reflecting memory-induced dynamics. Crucially, its self-attention mechanism preserves structure details and subtle correlations that the CNN loses.

This divergence highlights the importance of modeling path-dependent dynamics. The CNN's static multi-channel snapshots struggle to track
the subtle, memory-driven modulations of the phase-space distribution, particularly the slight shifts in fringe contrast that accumulate
over time. By contrast, the Swin Transformer, through its sensitivity to weak signals and adaptive conditioning on $\tau$, can extract these fine features from corrupted inputs and adjust reconstruction accordingly, maintaining clarity even when extrapolating beyond the training horizon.

These results indicate that for history-dependent environments, the ability to follow continuous trajectories and extract fine features
offers significant advantages, particularly when memory effects amplify subtle distortions at longer times. 


\section{Conclusion}
\label{sec:conclusion}

We have presented a neural architecture for extrapolative quantum error mitigation that overcomes the training horizon limitation inherent in existing machine-learning approaches. By integrating a Swin Transformer backbone with continuous time conditioning through adaptive layer normalization, the model learns generalizable correction functions that follow the dynamical accumulation of noise. This design enables accurate reconstruction of quantum states well beyond the temporal range covered by training data.

This synergistic approach addresses both the spatial and temporal challenges of long-time evolution. The architecture extracts weak structural correlations from degraded phase-space distributions, where environmental dissipation has suppressed fine quantum structures, while continuous time conditioning allows the model to capture the underlying law governing noise accumulation. Extensive numerical experiments across different Hamiltonians and noise regimes, including both Markovian and non-Markovian dynamics, demonstrate that the method maintains high reconstruction accuracy when extrapolating to twice the training horizon, successfully recovering quantum states in regimes where previous methods degrade. 

Looking forward, this framework may be extended to multimode continuous-variable systems~\cite{beckerClassicalShadowTomographyContinuousVariables2024,wuEfficientLearningContinuousvariablequantumstates2024,gandhariPrecisionBoundsContinuousVariableStateTomography2024,heEfficientMultimodeWignertomography2024} and further enhanced by incorporating physics-informed constraints into the learning process~\cite{hradilQuantumstateEstimation1997,rehacekDilutedMaximumlikelihoodAlgorithmquantumtomography2007,koutnyNeuralnetworkQuantumStatetomography2022,fengPhysicsInformedNeuralNetworksAdaptiveConstraints2025}. Overall, our results establish extrapolative quantum error mitigation as a practical strategy for reducing the experimental data requirements for faithful quantum-state reconstruction in continuous-variable platforms.


\begin{acknowledgments}
This work is supported by National Natural Science Foundation of China (Grant No. 12375025). 
\end{acknowledgments}
\bibliographystyle{apsrev4-2}

\begin{thebibliography}{77}%
\makeatletter
\providecommand \@ifxundefined [1]{%
 \@ifx{#1\undefined}
}%
\providecommand \@ifnum [1]{%
 \ifnum #1\expandafter \@firstoftwo
 \else \expandafter \@secondoftwo
 \fi
}%
\providecommand \@ifx [1]{%
 \ifx #1\expandafter \@firstoftwo
 \else \expandafter \@secondoftwo
 \fi
}%
\providecommand \natexlab [1]{#1}%
\providecommand \enquote  [1]{``#1''}%
\providecommand \bibnamefont  [1]{#1}%
\providecommand \bibfnamefont [1]{#1}%
\providecommand \citenamefont [1]{#1}%
\providecommand \href@noop [0]{\@secondoftwo}%
\providecommand \href [0]{\begingroup \@sanitize@url \@href}%
\providecommand \@href[1]{\@@startlink{#1}\@@href}%
\providecommand \@@href[1]{\endgroup#1\@@endlink}%
\providecommand \@sanitize@url [0]{\catcode `\\12\catcode `\$12\catcode `\&12\catcode `\#12\catcode `\^12\catcode `\_12\catcode `\%12\relax}%
\providecommand \@@startlink[1]{}%
\providecommand \@@endlink[0]{}%
\providecommand \url  [0]{\begingroup\@sanitize@url \@url }%
\providecommand \@url [1]{\endgroup\@href {#1}{\urlprefix }}%
\providecommand \urlprefix  [0]{URL }%
\providecommand \Eprint [0]{\href }%
\providecommand \doibase [0]{https://doi.org/}%
\providecommand \selectlanguage [0]{\@gobble}%
\providecommand \bibinfo  [0]{\@secondoftwo}%
\providecommand \bibfield  [0]{\@secondoftwo}%
\providecommand \translation [1]{[#1]}%
\providecommand \BibitemOpen [0]{}%
\providecommand \bibitemStop [0]{}%
\providecommand \bibitemNoStop [0]{.\EOS\space}%
\providecommand \EOS [0]{\spacefactor3000\relax}%
\providecommand \BibitemShut  [1]{\csname bibitem#1\endcsname}%
\let\auto@bib@innerbib\@empty
\bibitem [{\citenamefont {Abdo}\ \emph {et~al.}(2025)\citenamefont {Abdo}, \citenamefont {Shanks}, \citenamefont {Jinka}, \citenamefont {Rozen},\ and\ \citenamefont {Orcutt}}]{abdoTeleportationEntanglementSwappingContinuousQuantum2025}%
  \BibitemOpen
  \bibfield  {author} {\bibinfo {author} {\bibfnamefont {B.}~\bibnamefont {Abdo}}, \bibinfo {author} {\bibfnamefont {W.}~\bibnamefont {Shanks}}, \bibinfo {author} {\bibfnamefont {O.}~\bibnamefont {Jinka}}, \bibinfo {author} {\bibfnamefont {J.~R.}\ \bibnamefont {Rozen}},\ and\ \bibinfo {author} {\bibfnamefont {J.}~\bibnamefont {Orcutt}},\ }\href {https://doi.org/10.1103/9cpm-kr4h} {\bibfield  {journal} {\bibinfo  {journal} {Physical Review X}\ }\textbf {\bibinfo {volume} {15}},\ \bibinfo {pages} {031075} (\bibinfo {year} {2025})}\BibitemShut {NoStop}%
\bibitem [{\citenamefont {Lloyd}\ and\ \citenamefont {Braunstein}(1999)}]{lloydQuantumComputationContinuousVariables1999}%
  \BibitemOpen
  \bibfield  {author} {\bibinfo {author} {\bibfnamefont {S.}~\bibnamefont {Lloyd}}\ and\ \bibinfo {author} {\bibfnamefont {S.~L.}\ \bibnamefont {Braunstein}},\ }\href {https://doi.org/10.1103/PhysRevLett.82.1784} {\bibfield  {journal} {\bibinfo  {journal} {Physical Review Letters}\ }\textbf {\bibinfo {volume} {82}},\ \bibinfo {pages} {1784} (\bibinfo {year} {1999})}\BibitemShut {NoStop}%
\bibitem [{\citenamefont {Gottesman}\ \emph {et~al.}(2001)\citenamefont {Gottesman}, \citenamefont {Kitaev},\ and\ \citenamefont {Preskill}}]{gottesmanEncodingQubitOscillator2001}%
  \BibitemOpen
  \bibfield  {author} {\bibinfo {author} {\bibfnamefont {D.}~\bibnamefont {Gottesman}}, \bibinfo {author} {\bibfnamefont {A.}~\bibnamefont {Kitaev}},\ and\ \bibinfo {author} {\bibfnamefont {J.}~\bibnamefont {Preskill}},\ }\href {https://doi.org/10.1103/PhysRevA.64.012310} {\bibfield  {journal} {\bibinfo  {journal} {Physical Review A}\ }\textbf {\bibinfo {volume} {64}},\ \bibinfo {pages} {012310} (\bibinfo {year} {2001})}\BibitemShut {NoStop}%
\bibitem [{\citenamefont {Lvovsky}\ and\ \citenamefont {Raymer}(2009)}]{lvovskyContinuousvariableOpticalQuantumstatetomography2009}%
  \BibitemOpen
  \bibfield  {author} {\bibinfo {author} {\bibfnamefont {A.~I.}\ \bibnamefont {Lvovsky}}\ and\ \bibinfo {author} {\bibfnamefont {M.~G.}\ \bibnamefont {Raymer}},\ }\href {https://doi.org/10.1103/RevModPhys.81.299} {\bibfield  {journal} {\bibinfo  {journal} {Reviews of Modern Physics}\ }\textbf {\bibinfo {volume} {81}},\ \bibinfo {pages} {299} (\bibinfo {year} {2009})}\BibitemShut {NoStop}%
\bibitem [{\citenamefont {Adesso}\ \emph {et~al.}(2014)\citenamefont {Adesso}, \citenamefont {Ragy},\ and\ \citenamefont {Lee}}]{adessoContinuousVariableQuantumInformationGaussian2014}%
  \BibitemOpen
  \bibfield  {author} {\bibinfo {author} {\bibfnamefont {G.}~\bibnamefont {Adesso}}, \bibinfo {author} {\bibfnamefont {S.}~\bibnamefont {Ragy}},\ and\ \bibinfo {author} {\bibfnamefont {A.~R.}\ \bibnamefont {Lee}},\ }\href {https://doi.org/10.1142/S1230161214400010} {\bibfield  {journal} {\bibinfo  {journal} {Open Systems \& Information Dynamics}\ }\textbf {\bibinfo {volume} {21}},\ \bibinfo {pages} {1440001} (\bibinfo {year} {2014})}\BibitemShut {NoStop}%
\bibitem [{\citenamefont {Andersen}\ \emph {et~al.}(2015)\citenamefont {Andersen}, \citenamefont {{Neergaard-Nielsen}}, \citenamefont {Van~Loock},\ and\ \citenamefont {Furusawa}}]{andersenHybridDiscreteContinuousvariablequantuminformation2015}%
  \BibitemOpen
  \bibfield  {author} {\bibinfo {author} {\bibfnamefont {U.~L.}\ \bibnamefont {Andersen}}, \bibinfo {author} {\bibfnamefont {J.~S.}\ \bibnamefont {{Neergaard-Nielsen}}}, \bibinfo {author} {\bibfnamefont {P.}~\bibnamefont {Van~Loock}},\ and\ \bibinfo {author} {\bibfnamefont {A.}~\bibnamefont {Furusawa}},\ }\href {https://doi.org/10.1038/nphys3410} {\bibfield  {journal} {\bibinfo  {journal} {Nature Physics}\ }\textbf {\bibinfo {volume} {11}},\ \bibinfo {pages} {713} (\bibinfo {year} {2015})}\BibitemShut {NoStop}%
\bibitem [{\citenamefont {Yokoyama}\ \emph {et~al.}(2013)\citenamefont {Yokoyama}, \citenamefont {Ukai}, \citenamefont {Armstrong}, \citenamefont {Sornphiphatphong}, \citenamefont {Kaji}, \citenamefont {Suzuki}, \citenamefont {Yoshikawa}, \citenamefont {Yonezawa}, \citenamefont {Menicucci},\ and\ \citenamefont {Furusawa}}]{yokoyamaUltralargescaleContinuousvariableClusterstatesmultiplexed2013}%
  \BibitemOpen
  \bibfield  {author} {\bibinfo {author} {\bibfnamefont {S.}~\bibnamefont {Yokoyama}}, \bibinfo {author} {\bibfnamefont {R.}~\bibnamefont {Ukai}}, \bibinfo {author} {\bibfnamefont {S.~C.}\ \bibnamefont {Armstrong}}, \bibinfo {author} {\bibfnamefont {C.}~\bibnamefont {Sornphiphatphong}}, \bibinfo {author} {\bibfnamefont {T.}~\bibnamefont {Kaji}}, \bibinfo {author} {\bibfnamefont {S.}~\bibnamefont {Suzuki}}, \bibinfo {author} {\bibfnamefont {J.-i.}\ \bibnamefont {Yoshikawa}}, \bibinfo {author} {\bibfnamefont {H.}~\bibnamefont {Yonezawa}}, \bibinfo {author} {\bibfnamefont {N.~C.}\ \bibnamefont {Menicucci}},\ and\ \bibinfo {author} {\bibfnamefont {A.}~\bibnamefont {Furusawa}},\ }\href {https://doi.org/10.1038/nphoton.2013.287} {\bibfield  {journal} {\bibinfo  {journal} {Nature Photonics}\ }\textbf {\bibinfo {volume} {7}},\ \bibinfo {pages} {982} (\bibinfo {year} {2013})}\BibitemShut {NoStop}%
\bibitem [{\citenamefont {Asavanant}\ \emph {et~al.}(2019)\citenamefont {Asavanant}, \citenamefont {Shiozawa}, \citenamefont {Yokoyama}, \citenamefont {Charoensombutamon}, \citenamefont {Emura}, \citenamefont {Alexander}, \citenamefont {Takeda}, \citenamefont {Yoshikawa}, \citenamefont {Menicucci}, \citenamefont {Yonezawa},\ and\ \citenamefont {Furusawa}}]{asavanantGenerationTimedomainmultiplexedTwodimensionalclusterstate2019}%
  \BibitemOpen
  \bibfield  {author} {\bibinfo {author} {\bibfnamefont {W.}~\bibnamefont {Asavanant}}, \bibinfo {author} {\bibfnamefont {Y.}~\bibnamefont {Shiozawa}}, \bibinfo {author} {\bibfnamefont {S.}~\bibnamefont {Yokoyama}}, \bibinfo {author} {\bibfnamefont {B.}~\bibnamefont {Charoensombutamon}}, \bibinfo {author} {\bibfnamefont {H.}~\bibnamefont {Emura}}, \bibinfo {author} {\bibfnamefont {R.~N.}\ \bibnamefont {Alexander}}, \bibinfo {author} {\bibfnamefont {S.}~\bibnamefont {Takeda}}, \bibinfo {author} {\bibfnamefont {J.-i.}\ \bibnamefont {Yoshikawa}}, \bibinfo {author} {\bibfnamefont {N.~C.}\ \bibnamefont {Menicucci}}, \bibinfo {author} {\bibfnamefont {H.}~\bibnamefont {Yonezawa}},\ and\ \bibinfo {author} {\bibfnamefont {A.}~\bibnamefont {Furusawa}},\ }\href {https://doi.org/10.1126/science.aay2645} {\bibfield  {journal} {\bibinfo  {journal} {Science}\ }\textbf {\bibinfo {volume} {366}},\ \bibinfo {pages} {373} (\bibinfo {year} {2019})}\BibitemShut {NoStop}%
\bibitem [{\citenamefont {Grosshans}\ and\ \citenamefont {Grangier}(2002)}]{grosshansContinuousVariableQuantumCryptographyUsing2002}%
  \BibitemOpen
  \bibfield  {author} {\bibinfo {author} {\bibfnamefont {F.}~\bibnamefont {Grosshans}}\ and\ \bibinfo {author} {\bibfnamefont {P.}~\bibnamefont {Grangier}},\ }\href {https://doi.org/10.1103/PhysRevLett.88.057902} {\bibfield  {journal} {\bibinfo  {journal} {Physical Review Letters}\ }\textbf {\bibinfo {volume} {88}},\ \bibinfo {pages} {057902} (\bibinfo {year} {2002})}\BibitemShut {NoStop}%
\bibitem [{\citenamefont {Gu}\ \emph {et~al.}(2009)\citenamefont {Gu}, \citenamefont {Weedbrook}, \citenamefont {Menicucci}, \citenamefont {Ralph},\ and\ \citenamefont {Van~Loock}}]{guQuantumComputingContinuousvariableclusters2009}%
  \BibitemOpen
  \bibfield  {author} {\bibinfo {author} {\bibfnamefont {M.}~\bibnamefont {Gu}}, \bibinfo {author} {\bibfnamefont {C.}~\bibnamefont {Weedbrook}}, \bibinfo {author} {\bibfnamefont {N.~C.}\ \bibnamefont {Menicucci}}, \bibinfo {author} {\bibfnamefont {T.~C.}\ \bibnamefont {Ralph}},\ and\ \bibinfo {author} {\bibfnamefont {P.}~\bibnamefont {Van~Loock}},\ }\href {https://doi.org/10.1103/PhysRevA.79.062318} {\bibfield  {journal} {\bibinfo  {journal} {Physical Review A}\ }\textbf {\bibinfo {volume} {79}},\ \bibinfo {pages} {062318} (\bibinfo {year} {2009})}\BibitemShut {NoStop}%
\bibitem [{\citenamefont {Pirandola}\ \emph {et~al.}(2020)\citenamefont {Pirandola}, \citenamefont {Andersen}, \citenamefont {Banchi}, \citenamefont {Berta}, \citenamefont {Bunandar}, \citenamefont {Colbeck}, \citenamefont {Englund}, \citenamefont {Gehring}, \citenamefont {Lupo}, \citenamefont {Ottaviani}, \citenamefont {Pereira}, \citenamefont {Razavi}, \citenamefont {Shamsul~Shaari}, \citenamefont {Tomamichel}, \citenamefont {Usenko}, \citenamefont {Vallone}, \citenamefont {Villoresi},\ and\ \citenamefont {Wallden}}]{pirandolaAdvancesQuantumCryptography2020}%
  \BibitemOpen
  \bibfield  {author} {\bibinfo {author} {\bibfnamefont {S.}~\bibnamefont {Pirandola}}, \bibinfo {author} {\bibfnamefont {U.~L.}\ \bibnamefont {Andersen}}, \bibinfo {author} {\bibfnamefont {L.}~\bibnamefont {Banchi}}, \bibinfo {author} {\bibfnamefont {M.}~\bibnamefont {Berta}}, \bibinfo {author} {\bibfnamefont {D.}~\bibnamefont {Bunandar}}, \bibinfo {author} {\bibfnamefont {R.}~\bibnamefont {Colbeck}}, \bibinfo {author} {\bibfnamefont {D.}~\bibnamefont {Englund}}, \bibinfo {author} {\bibfnamefont {T.}~\bibnamefont {Gehring}}, \bibinfo {author} {\bibfnamefont {C.}~\bibnamefont {Lupo}}, \bibinfo {author} {\bibfnamefont {C.}~\bibnamefont {Ottaviani}}, \bibinfo {author} {\bibfnamefont {J.~L.}\ \bibnamefont {Pereira}}, \bibinfo {author} {\bibfnamefont {M.}~\bibnamefont {Razavi}}, \bibinfo {author} {\bibfnamefont {J.}~\bibnamefont {Shamsul~Shaari}}, \bibinfo {author} {\bibfnamefont {M.}~\bibnamefont {Tomamichel}}, \bibinfo {author} {\bibfnamefont {V.~C.}\ \bibnamefont {Usenko}}, \bibinfo {author} {\bibfnamefont
  {G.}~\bibnamefont {Vallone}}, \bibinfo {author} {\bibfnamefont {P.}~\bibnamefont {Villoresi}},\ and\ \bibinfo {author} {\bibfnamefont {P.}~\bibnamefont {Wallden}},\ }\href {https://doi.org/10.1364/AOP.361502} {\bibfield  {journal} {\bibinfo  {journal} {Advances in Optics and Photonics}\ }\textbf {\bibinfo {volume} {12}},\ \bibinfo {pages} {1012} (\bibinfo {year} {2020})}\BibitemShut {NoStop}%
\bibitem [{\citenamefont {Pirandola}\ \emph {et~al.}(2017)\citenamefont {Pirandola}, \citenamefont {Laurenza}, \citenamefont {Ottaviani},\ and\ \citenamefont {Banchi}}]{pirandolaFundamentalLimitsRepeaterlessquantumcommunications2017}%
  \BibitemOpen
  \bibfield  {author} {\bibinfo {author} {\bibfnamefont {S.}~\bibnamefont {Pirandola}}, \bibinfo {author} {\bibfnamefont {R.}~\bibnamefont {Laurenza}}, \bibinfo {author} {\bibfnamefont {C.}~\bibnamefont {Ottaviani}},\ and\ \bibinfo {author} {\bibfnamefont {L.}~\bibnamefont {Banchi}},\ }\href {https://doi.org/10.1038/ncomms15043} {\bibfield  {journal} {\bibinfo  {journal} {Nature Communications}\ }\textbf {\bibinfo {volume} {8}},\ \bibinfo {pages} {15043} (\bibinfo {year} {2017})}\BibitemShut {NoStop}%
\bibitem [{\citenamefont {Takeoka}\ \emph {et~al.}(2014)\citenamefont {Takeoka}, \citenamefont {Guha},\ and\ \citenamefont {Wilde}}]{takeokaFundamentalRatelossTradeoffopticalquantum2014}%
  \BibitemOpen
  \bibfield  {author} {\bibinfo {author} {\bibfnamefont {M.}~\bibnamefont {Takeoka}}, \bibinfo {author} {\bibfnamefont {S.}~\bibnamefont {Guha}},\ and\ \bibinfo {author} {\bibfnamefont {M.~M.}\ \bibnamefont {Wilde}},\ }\href {https://doi.org/10.1038/ncomms6235} {\bibfield  {journal} {\bibinfo  {journal} {Nature Communications}\ }\textbf {\bibinfo {volume} {5}},\ \bibinfo {pages} {5235} (\bibinfo {year} {2014})}\BibitemShut {NoStop}%
\bibitem [{\citenamefont {Holevo}\ and\ \citenamefont {Werner}(2001)}]{holevoEvaluatingCapacitiesBosonicGaussianchannels2001}%
  \BibitemOpen
  \bibfield  {author} {\bibinfo {author} {\bibfnamefont {A.}~\bibnamefont {Holevo}}\ and\ \bibinfo {author} {\bibfnamefont {R.}~\bibnamefont {Werner}},\ }\href {https://doi.org/10.1103/PhysRevA.63.032312} {\bibfield  {journal} {\bibinfo  {journal} {Physical Review A}\ }\textbf {\bibinfo {volume} {63}},\ \bibinfo {pages} {032312} (\bibinfo {year} {2001})}\BibitemShut {NoStop}%
\bibitem [{\citenamefont {Michael}\ \emph {et~al.}(2016)\citenamefont {Michael}, \citenamefont {Silveri}, \citenamefont {Brierley}, \citenamefont {Albert}, \citenamefont {Salmilehto}, \citenamefont {Jiang},\ and\ \citenamefont {Girvin}}]{michaelNewClassQuantumErrorCorrectingCodes2016}%
  \BibitemOpen
  \bibfield  {author} {\bibinfo {author} {\bibfnamefont {M.~H.}\ \bibnamefont {Michael}}, \bibinfo {author} {\bibfnamefont {M.}~\bibnamefont {Silveri}}, \bibinfo {author} {\bibfnamefont {R.~T.}\ \bibnamefont {Brierley}}, \bibinfo {author} {\bibfnamefont {V.~V.}\ \bibnamefont {Albert}}, \bibinfo {author} {\bibfnamefont {J.}~\bibnamefont {Salmilehto}}, \bibinfo {author} {\bibfnamefont {L.}~\bibnamefont {Jiang}},\ and\ \bibinfo {author} {\bibfnamefont {S.~M.}\ \bibnamefont {Girvin}},\ }\href {https://doi.org/10.1103/PhysRevX.6.031006} {\bibfield  {journal} {\bibinfo  {journal} {Physical Review X}\ }\textbf {\bibinfo {volume} {6}},\ \bibinfo {pages} {031006} (\bibinfo {year} {2016})}\BibitemShut {NoStop}%
\bibitem [{\citenamefont {Leghtas}\ \emph {et~al.}(2013)\citenamefont {Leghtas}, \citenamefont {Kirchmair}, \citenamefont {Vlastakis}, \citenamefont {Schoelkopf}, \citenamefont {Devoret},\ and\ \citenamefont {Mirrahimi}}]{leghtasHardwareEfficientAutonomousQuantumMemoryProtection2013}%
  \BibitemOpen
  \bibfield  {author} {\bibinfo {author} {\bibfnamefont {Z.}~\bibnamefont {Leghtas}}, \bibinfo {author} {\bibfnamefont {G.}~\bibnamefont {Kirchmair}}, \bibinfo {author} {\bibfnamefont {B.}~\bibnamefont {Vlastakis}}, \bibinfo {author} {\bibfnamefont {R.~J.}\ \bibnamefont {Schoelkopf}}, \bibinfo {author} {\bibfnamefont {M.~H.}\ \bibnamefont {Devoret}},\ and\ \bibinfo {author} {\bibfnamefont {M.}~\bibnamefont {Mirrahimi}},\ }\href {https://doi.org/10.1103/PhysRevLett.111.120501} {\bibfield  {journal} {\bibinfo  {journal} {Physical Review Letters}\ }\textbf {\bibinfo {volume} {111}},\ \bibinfo {pages} {120501} (\bibinfo {year} {2013})}\BibitemShut {NoStop}%
\bibitem [{\citenamefont {Mirrahimi}\ \emph {et~al.}(2014)\citenamefont {Mirrahimi}, \citenamefont {Leghtas}, \citenamefont {Albert}, \citenamefont {Touzard}, \citenamefont {Schoelkopf}, \citenamefont {Jiang},\ and\ \citenamefont {Devoret}}]{mirrahimiDynamicallyProtectedCatqubitsnewparadigm2014}%
  \BibitemOpen
  \bibfield  {author} {\bibinfo {author} {\bibfnamefont {M.}~\bibnamefont {Mirrahimi}}, \bibinfo {author} {\bibfnamefont {Z.}~\bibnamefont {Leghtas}}, \bibinfo {author} {\bibfnamefont {V.~V.}\ \bibnamefont {Albert}}, \bibinfo {author} {\bibfnamefont {S.}~\bibnamefont {Touzard}}, \bibinfo {author} {\bibfnamefont {R.~J.}\ \bibnamefont {Schoelkopf}}, \bibinfo {author} {\bibfnamefont {L.}~\bibnamefont {Jiang}},\ and\ \bibinfo {author} {\bibfnamefont {M.~H.}\ \bibnamefont {Devoret}},\ }\href {https://doi.org/10.1088/1367-2630/16/4/045014} {\bibfield  {journal} {\bibinfo  {journal} {New Journal of Physics}\ }\textbf {\bibinfo {volume} {16}},\ \bibinfo {pages} {045014} (\bibinfo {year} {2014})}\BibitemShut {NoStop}%
\bibitem [{\citenamefont {Ofek}\ \emph {et~al.}(2016)\citenamefont {Ofek}, \citenamefont {Petrenko}, \citenamefont {Heeres}, \citenamefont {Reinhold}, \citenamefont {Leghtas}, \citenamefont {Vlastakis}, \citenamefont {Liu}, \citenamefont {Frunzio}, \citenamefont {Girvin}, \citenamefont {Jiang}, \citenamefont {Mirrahimi}, \citenamefont {Devoret},\ and\ \citenamefont {Schoelkopf}}]{ofekExtendingLifetimeQuantumbiterror2016a}%
  \BibitemOpen
  \bibfield  {author} {\bibinfo {author} {\bibfnamefont {N.}~\bibnamefont {Ofek}}, \bibinfo {author} {\bibfnamefont {A.}~\bibnamefont {Petrenko}}, \bibinfo {author} {\bibfnamefont {R.}~\bibnamefont {Heeres}}, \bibinfo {author} {\bibfnamefont {P.}~\bibnamefont {Reinhold}}, \bibinfo {author} {\bibfnamefont {Z.}~\bibnamefont {Leghtas}}, \bibinfo {author} {\bibfnamefont {B.}~\bibnamefont {Vlastakis}}, \bibinfo {author} {\bibfnamefont {Y.}~\bibnamefont {Liu}}, \bibinfo {author} {\bibfnamefont {L.}~\bibnamefont {Frunzio}}, \bibinfo {author} {\bibfnamefont {S.~M.}\ \bibnamefont {Girvin}}, \bibinfo {author} {\bibfnamefont {L.}~\bibnamefont {Jiang}}, \bibinfo {author} {\bibfnamefont {M.}~\bibnamefont {Mirrahimi}}, \bibinfo {author} {\bibfnamefont {M.~H.}\ \bibnamefont {Devoret}},\ and\ \bibinfo {author} {\bibfnamefont {R.~J.}\ \bibnamefont {Schoelkopf}},\ }\href {https://doi.org/10.1038/nature18949} {\bibfield  {journal} {\bibinfo  {journal} {Nature}\ }\textbf {\bibinfo {volume} {536}},\ \bibinfo {pages} {441}
  (\bibinfo {year} {2016})}\BibitemShut {NoStop}%
\bibitem [{\citenamefont {Vasconcelos}\ \emph {et~al.}(2010)\citenamefont {Vasconcelos}, \citenamefont {Sanz},\ and\ \citenamefont {Glancy}}]{vasconcelosAllopticalGenerationStatesEncodingqubit2010}%
  \BibitemOpen
  \bibfield  {author} {\bibinfo {author} {\bibfnamefont {H.~M.}\ \bibnamefont {Vasconcelos}}, \bibinfo {author} {\bibfnamefont {L.}~\bibnamefont {Sanz}},\ and\ \bibinfo {author} {\bibfnamefont {S.}~\bibnamefont {Glancy}},\ }\href {https://doi.org/10.1364/OL.35.003261} {\bibfield  {journal} {\bibinfo  {journal} {Optics Letters}\ }\textbf {\bibinfo {volume} {35}},\ \bibinfo {pages} {3261} (\bibinfo {year} {2010})}\BibitemShut {NoStop}%
\bibitem [{\citenamefont {Andersen}\ \emph {et~al.}(2010)\citenamefont {Andersen}, \citenamefont {Leuchs},\ and\ \citenamefont {Silberhorn}}]{andersenContinuousvariableQuantumInformationprocessing2010}%
  \BibitemOpen
  \bibfield  {author} {\bibinfo {author} {\bibfnamefont {U.}~\bibnamefont {Andersen}}, \bibinfo {author} {\bibfnamefont {G.}~\bibnamefont {Leuchs}},\ and\ \bibinfo {author} {\bibfnamefont {C.}~\bibnamefont {Silberhorn}},\ }\href {https://doi.org/10.1002/lpor.200910010} {\bibfield  {journal} {\bibinfo  {journal} {Laser \& Photonics Reviews}\ }\textbf {\bibinfo {volume} {4}},\ \bibinfo {pages} {337} (\bibinfo {year} {2010})}\BibitemShut {NoStop}%
\bibitem [{\citenamefont {Arzani}\ \emph {et~al.}(2025)\citenamefont {Arzani}, \citenamefont {Booth},\ and\ \citenamefont {Chabaud}}]{arzaniEffectiveDescriptionsBosonicsystemscan2025}%
  \BibitemOpen
  \bibfield  {author} {\bibinfo {author} {\bibfnamefont {F.}~\bibnamefont {Arzani}}, \bibinfo {author} {\bibfnamefont {R.~I.}\ \bibnamefont {Booth}},\ and\ \bibinfo {author} {\bibfnamefont {U.}~\bibnamefont {Chabaud}},\ }\href {https://doi.org/10.1038/s41467-025-64872-3} {\bibfield  {journal} {\bibinfo  {journal} {Nature Communications}\ }\textbf {\bibinfo {volume} {16}},\ \bibinfo {pages} {9744} (\bibinfo {year} {2025})}\BibitemShut {NoStop}%
\bibitem [{\citenamefont {Braunstein}\ and\ \citenamefont {Van~Loock}(2005)}]{braunsteinQuantumInformationContinuousvariables2005}%
  \BibitemOpen
  \bibfield  {author} {\bibinfo {author} {\bibfnamefont {S.~L.}\ \bibnamefont {Braunstein}}\ and\ \bibinfo {author} {\bibfnamefont {P.}~\bibnamefont {Van~Loock}},\ }\href {https://doi.org/10.1103/RevModPhys.77.513} {\bibfield  {journal} {\bibinfo  {journal} {Reviews of Modern Physics}\ }\textbf {\bibinfo {volume} {77}},\ \bibinfo {pages} {513} (\bibinfo {year} {2005})}\BibitemShut {NoStop}%
\bibitem [{\citenamefont {Larsen}\ \emph {et~al.}(2021)\citenamefont {Larsen}, \citenamefont {Chamberland}, \citenamefont {Noh}, \citenamefont {{Neergaard-Nielsen}},\ and\ \citenamefont {Andersen}}]{larsenFaultTolerantContinuousVariableMeasurementbasedQuantumComputation2021}%
  \BibitemOpen
  \bibfield  {author} {\bibinfo {author} {\bibfnamefont {M.~V.}\ \bibnamefont {Larsen}}, \bibinfo {author} {\bibfnamefont {C.}~\bibnamefont {Chamberland}}, \bibinfo {author} {\bibfnamefont {K.}~\bibnamefont {Noh}}, \bibinfo {author} {\bibfnamefont {J.~S.}\ \bibnamefont {{Neergaard-Nielsen}}},\ and\ \bibinfo {author} {\bibfnamefont {U.~L.}\ \bibnamefont {Andersen}},\ }\href {https://doi.org/10.1103/PRXQuantum.2.030325} {\bibfield  {journal} {\bibinfo  {journal} {PRX Quantum}\ }\textbf {\bibinfo {volume} {2}},\ \bibinfo {pages} {030325} (\bibinfo {year} {2021})}\BibitemShut {NoStop}%
\bibitem [{\citenamefont {Ma}\ \emph {et~al.}(2021)\citenamefont {Ma}, \citenamefont {Puri}, \citenamefont {Schoelkopf}, \citenamefont {Devoret}, \citenamefont {Girvin},\ and\ \citenamefont {Jiang}}]{maQuantumControlBosonicmodessuperconducting2021}%
  \BibitemOpen
  \bibfield  {author} {\bibinfo {author} {\bibfnamefont {W.-L.}\ \bibnamefont {Ma}}, \bibinfo {author} {\bibfnamefont {S.}~\bibnamefont {Puri}}, \bibinfo {author} {\bibfnamefont {R.~J.}\ \bibnamefont {Schoelkopf}}, \bibinfo {author} {\bibfnamefont {M.~H.}\ \bibnamefont {Devoret}}, \bibinfo {author} {\bibfnamefont {S.}~\bibnamefont {Girvin}},\ and\ \bibinfo {author} {\bibfnamefont {L.}~\bibnamefont {Jiang}},\ }\href {https://doi.org/10.1016/j.scib.2021.05.024} {\bibfield  {journal} {\bibinfo  {journal} {Science Bulletin}\ }\textbf {\bibinfo {volume} {66}},\ \bibinfo {pages} {1789} (\bibinfo {year} {2021})}\BibitemShut {NoStop}%
\bibitem [{\citenamefont {Nielsen}\ and\ \citenamefont {Chuang}(2012)}]{nielsenQuantumComputationQuantum2012}%
  \BibitemOpen
  \bibfield  {author} {\bibinfo {author} {\bibfnamefont {M.~A.}\ \bibnamefont {Nielsen}}\ and\ \bibinfo {author} {\bibfnamefont {I.~L.}\ \bibnamefont {Chuang}},\ }\href {https://doi.org/10.1017/CBO9780511976667} {\emph {\bibinfo {title} {Quantum Computation and Quantum Information: 10th Anniversary Edition}}},\ \bibinfo {edition} {1st}\ ed.\ (\bibinfo  {publisher} {Cambridge University Press},\ \bibinfo {year} {2012})\BibitemShut {NoStop}%
\bibitem [{\citenamefont {Weedbrook}\ \emph {et~al.}(2012)\citenamefont {Weedbrook}, \citenamefont {Pirandola}, \citenamefont {{Garc{\'i}a-Patr{\'o}n}}, \citenamefont {Cerf}, \citenamefont {Ralph}, \citenamefont {Shapiro},\ and\ \citenamefont {Lloyd}}]{weedbrookGaussianQuantumInformation2012a}%
  \BibitemOpen
  \bibfield  {author} {\bibinfo {author} {\bibfnamefont {C.}~\bibnamefont {Weedbrook}}, \bibinfo {author} {\bibfnamefont {S.}~\bibnamefont {Pirandola}}, \bibinfo {author} {\bibfnamefont {R.}~\bibnamefont {{Garc{\'i}a-Patr{\'o}n}}}, \bibinfo {author} {\bibfnamefont {N.~J.}\ \bibnamefont {Cerf}}, \bibinfo {author} {\bibfnamefont {T.~C.}\ \bibnamefont {Ralph}}, \bibinfo {author} {\bibfnamefont {J.~H.}\ \bibnamefont {Shapiro}},\ and\ \bibinfo {author} {\bibfnamefont {S.}~\bibnamefont {Lloyd}},\ }\href {https://doi.org/10.1103/RevModPhys.84.621} {\bibfield  {journal} {\bibinfo  {journal} {Reviews of Modern Physics}\ }\textbf {\bibinfo {volume} {84}},\ \bibinfo {pages} {621} (\bibinfo {year} {2012})}\BibitemShut {NoStop}%
\bibitem [{\citenamefont {Gravina}\ \emph {et~al.}()\citenamefont {Gravina}, \citenamefont {Minganti},\ and\ \citenamefont {Savona}}]{gravinaCriticalSchrodingerCat2023}%
  \BibitemOpen
  \bibfield  {author} {\bibinfo {author} {\bibfnamefont {L.}~\bibnamefont {Gravina}}, \bibinfo {author} {\bibfnamefont {F.}~\bibnamefont {Minganti}},\ and\ \bibinfo {author} {\bibfnamefont {V.}~\bibnamefont {Savona}},\ }\href {https://doi.org/10.1103/PRXQuantum.4.020337} {\bibfield  {journal} {\bibinfo  {journal} {PRX Quantum}\ }\textbf {\bibinfo {volume} {4}},\ \bibinfo {pages} {020337}}\BibitemShut {NoStop}%
\bibitem [{\citenamefont {Cai}\ \emph {et~al.}(2023)\citenamefont {Cai}, \citenamefont {Babbush}, \citenamefont {Benjamin}, \citenamefont {Endo}, \citenamefont {Huggins}, \citenamefont {Li}, \citenamefont {McClean},\ and\ \citenamefont {O'Brien}}]{caiQuantumErrorMitigation2023}%
  \BibitemOpen
  \bibfield  {author} {\bibinfo {author} {\bibfnamefont {Z.}~\bibnamefont {Cai}}, \bibinfo {author} {\bibfnamefont {R.}~\bibnamefont {Babbush}}, \bibinfo {author} {\bibfnamefont {S.~C.}\ \bibnamefont {Benjamin}}, \bibinfo {author} {\bibfnamefont {S.}~\bibnamefont {Endo}}, \bibinfo {author} {\bibfnamefont {W.~J.}\ \bibnamefont {Huggins}}, \bibinfo {author} {\bibfnamefont {Y.}~\bibnamefont {Li}}, \bibinfo {author} {\bibfnamefont {J.~R.}\ \bibnamefont {McClean}},\ and\ \bibinfo {author} {\bibfnamefont {T.~E.}\ \bibnamefont {O'Brien}},\ }\href {https://doi.org/10.1103/RevModPhys.95.045005} {\bibfield  {journal} {\bibinfo  {journal} {Reviews of Modern Physics}\ }\textbf {\bibinfo {volume} {95}},\ \bibinfo {pages} {045005} (\bibinfo {year} {2023})}\BibitemShut {NoStop}%
\bibitem [{\citenamefont {{Giurgica-Tiron}}\ \emph {et~al.}(2020)\citenamefont {{Giurgica-Tiron}}, \citenamefont {Hindy}, \citenamefont {LaRose}, \citenamefont {Mari},\ and\ \citenamefont {Zeng}}]{giurgica-tironDigitalZeroNoise2020}%
  \BibitemOpen
  \bibfield  {author} {\bibinfo {author} {\bibfnamefont {T.}~\bibnamefont {{Giurgica-Tiron}}}, \bibinfo {author} {\bibfnamefont {Y.}~\bibnamefont {Hindy}}, \bibinfo {author} {\bibfnamefont {R.}~\bibnamefont {LaRose}}, \bibinfo {author} {\bibfnamefont {A.}~\bibnamefont {Mari}},\ and\ \bibinfo {author} {\bibfnamefont {W.~J.}\ \bibnamefont {Zeng}},\ }in\ \href {https://doi.org/10.1109/QCE49297.2020.00045} {\emph {\bibinfo {booktitle} {2020 IEEE International Conference on Quantum Computing and Engineering (QCE)}}}\ (\bibinfo  {publisher} {IEEE},\ \bibinfo {address} {Denver, CO, USA},\ \bibinfo {year} {2020})\ pp.\ \bibinfo {pages} {306--316}\BibitemShut {NoStop}%
\bibitem [{\citenamefont {Van Den~Berg}\ \emph {et~al.}(2023)\citenamefont {Van Den~Berg}, \citenamefont {Minev}, \citenamefont {Kandala},\ and\ \citenamefont {Temme}}]{vandenbergProbabilisticErrorCancellation2023}%
  \BibitemOpen
  \bibfield  {author} {\bibinfo {author} {\bibfnamefont {E.}~\bibnamefont {Van Den~Berg}}, \bibinfo {author} {\bibfnamefont {Z.~K.}\ \bibnamefont {Minev}}, \bibinfo {author} {\bibfnamefont {A.}~\bibnamefont {Kandala}},\ and\ \bibinfo {author} {\bibfnamefont {K.}~\bibnamefont {Temme}},\ }\href {https://doi.org/10.1038/s41567-023-02042-2} {\bibfield  {journal} {\bibinfo  {journal} {Nature Physics}\ }\textbf {\bibinfo {volume} {19}},\ \bibinfo {pages} {1116} (\bibinfo {year} {2023})}\BibitemShut {NoStop}%
\bibitem [{\citenamefont {McArdle}\ \emph {et~al.}(2021)\citenamefont {McArdle}, \citenamefont {O'Brien}, \citenamefont {Lee}, \citenamefont {Rubin}, \citenamefont {Boixo}, \citenamefont {Whaley}, \citenamefont {Babbush},\ and\ \citenamefont {McClean}}]{hugginsVirtualDistillationQuantum2021}%
  \BibitemOpen
  \bibfield  {author} {\bibinfo {author} {\bibfnamefont {S.}~\bibnamefont {McArdle}}, \bibinfo {author} {\bibfnamefont {T.~E.}\ \bibnamefont {O'Brien}}, \bibinfo {author} {\bibfnamefont {J.}~\bibnamefont {Lee}}, \bibinfo {author} {\bibfnamefont {N.~C.}\ \bibnamefont {Rubin}}, \bibinfo {author} {\bibfnamefont {S.}~\bibnamefont {Boixo}}, \bibinfo {author} {\bibfnamefont {K.~B.}\ \bibnamefont {Whaley}}, \bibinfo {author} {\bibfnamefont {R.}~\bibnamefont {Babbush}},\ and\ \bibinfo {author} {\bibfnamefont {J.~R.}\ \bibnamefont {McClean}},\ }\href {https://doi.org/10.1103/PhysRevX.11.041036} {\bibfield  {journal} {\bibinfo  {journal} {Physical Review X}\ }\textbf {\bibinfo {volume} {11}},\ \bibinfo {pages} {041036} (\bibinfo {year} {2021})}\BibitemShut {NoStop}%
\bibitem [{\citenamefont {Lowe}\ \emph {et~al.}(2021)\citenamefont {Lowe}, \citenamefont {Gordon}, \citenamefont {Czarnik}, \citenamefont {Arrasmith}, \citenamefont {Coles},\ and\ \citenamefont {Cincio}}]{loweUnifiedApproachDatadriven2021}%
  \BibitemOpen
  \bibfield  {author} {\bibinfo {author} {\bibfnamefont {A.}~\bibnamefont {Lowe}}, \bibinfo {author} {\bibfnamefont {M.~H.}\ \bibnamefont {Gordon}}, \bibinfo {author} {\bibfnamefont {P.}~\bibnamefont {Czarnik}}, \bibinfo {author} {\bibfnamefont {A.}~\bibnamefont {Arrasmith}}, \bibinfo {author} {\bibfnamefont {P.~J.}\ \bibnamefont {Coles}},\ and\ \bibinfo {author} {\bibfnamefont {L.}~\bibnamefont {Cincio}},\ }\href {https://doi.org/10.1103/PhysRevResearch.3.033098} {\bibfield  {journal} {\bibinfo  {journal} {Physical Review Research}\ }\textbf {\bibinfo {volume} {3}},\ \bibinfo {pages} {033098} (\bibinfo {year} {2021})}\BibitemShut {NoStop}%
\bibitem [{\citenamefont {Serafini}\ \emph {et~al.}(2005)\citenamefont {Serafini}, \citenamefont {Paris}, \citenamefont {Illuminati},\ and\ \citenamefont {Siena}}]{serafiniQuantifyingDecoherenceContinuousvariablesystems2005}%
  \BibitemOpen
  \bibfield  {author} {\bibinfo {author} {\bibfnamefont {A.}~\bibnamefont {Serafini}}, \bibinfo {author} {\bibfnamefont {M.~G.~A.}\ \bibnamefont {Paris}}, \bibinfo {author} {\bibfnamefont {F.}~\bibnamefont {Illuminati}},\ and\ \bibinfo {author} {\bibfnamefont {S.~D.}\ \bibnamefont {Siena}},\ }\href {https://doi.org/10.1088/1464-4266/7/4/R01} {\bibfield  {journal} {\bibinfo  {journal} {Journal of Optics B: Quantum and Semiclassical Optics}\ }\textbf {\bibinfo {volume} {7}},\ \bibinfo {pages} {R19} (\bibinfo {year} {2005})}\BibitemShut {NoStop}%
\bibitem [{\citenamefont {Walls}\ and\ \citenamefont {Milburn}(1994)}]{wallsQuantumOptics1994}%
  \BibitemOpen
  \bibfield  {author} {\bibinfo {author} {\bibfnamefont {D.~F.}\ \bibnamefont {Walls}}\ and\ \bibinfo {author} {\bibfnamefont {G.~J.}\ \bibnamefont {Milburn}},\ }\href {https://doi.org/10.1007/978-3-642-79504-6} {\emph {\bibinfo {title} {Quantum {{Optics}}}}}\ (\bibinfo  {publisher} {Springer Berlin Heidelberg},\ \bibinfo {address} {Berlin, Heidelberg},\ \bibinfo {year} {1994})\BibitemShut {NoStop}%
\bibitem [{\citenamefont {Leviant}\ \emph {et~al.}(2022)\citenamefont {Leviant}, \citenamefont {Xu}, \citenamefont {Jiang},\ and\ \citenamefont {Rosenblum}}]{leviantQuantumCapacityCodesbosoniclossdephasing2022}%
  \BibitemOpen
  \bibfield  {author} {\bibinfo {author} {\bibfnamefont {P.}~\bibnamefont {Leviant}}, \bibinfo {author} {\bibfnamefont {Q.}~\bibnamefont {Xu}}, \bibinfo {author} {\bibfnamefont {L.}~\bibnamefont {Jiang}},\ and\ \bibinfo {author} {\bibfnamefont {S.}~\bibnamefont {Rosenblum}},\ }\href {https://doi.org/10.22331/q-2022-09-29-821} {\bibfield  {journal} {\bibinfo  {journal} {Quantum}\ }\textbf {\bibinfo {volume} {6}},\ \bibinfo {pages} {821} (\bibinfo {year} {2022})},\ \Eprint {https://arxiv.org/abs/2205.00341} {arXiv:2205.00341 [quant-ph]} \BibitemShut {NoStop}%
\bibitem [{\citenamefont {Mele}\ \emph {et~al.}(2024)\citenamefont {Mele}, \citenamefont {Salek}, \citenamefont {Giovannetti},\ and\ \citenamefont {Lami}}]{meleQuantumCommunicationBosoniclossdephasingchannel2024}%
  \BibitemOpen
  \bibfield  {author} {\bibinfo {author} {\bibfnamefont {F.~A.}\ \bibnamefont {Mele}}, \bibinfo {author} {\bibfnamefont {F.}~\bibnamefont {Salek}}, \bibinfo {author} {\bibfnamefont {V.}~\bibnamefont {Giovannetti}},\ and\ \bibinfo {author} {\bibfnamefont {L.}~\bibnamefont {Lami}},\ }\href {https://doi.org/10.1103/PhysRevA.110.012460} {\bibfield  {journal} {\bibinfo  {journal} {Physical Review A}\ }\textbf {\bibinfo {volume} {110}},\ \bibinfo {pages} {012460} (\bibinfo {year} {2024})}\BibitemShut {NoStop}%
\bibitem [{\citenamefont {Wang}\ \emph {et~al.}(2009)\citenamefont {Wang}, \citenamefont {Hofheinz}, \citenamefont {Ansmann}, \citenamefont {Bialczak}, \citenamefont {Lucero}, \citenamefont {Neeley}, \citenamefont {O'Connell}, \citenamefont {Sank}, \citenamefont {Weides}, \citenamefont {Wenner}, \citenamefont {Cleland},\ and\ \citenamefont {Martinis}}]{wangDecoherenceDynamicsComplexPhotonStates2009}%
  \BibitemOpen
  \bibfield  {author} {\bibinfo {author} {\bibfnamefont {H.}~\bibnamefont {Wang}}, \bibinfo {author} {\bibfnamefont {M.}~\bibnamefont {Hofheinz}}, \bibinfo {author} {\bibfnamefont {M.}~\bibnamefont {Ansmann}}, \bibinfo {author} {\bibfnamefont {R.~C.}\ \bibnamefont {Bialczak}}, \bibinfo {author} {\bibfnamefont {E.}~\bibnamefont {Lucero}}, \bibinfo {author} {\bibfnamefont {M.}~\bibnamefont {Neeley}}, \bibinfo {author} {\bibfnamefont {A.~D.}\ \bibnamefont {O'Connell}}, \bibinfo {author} {\bibfnamefont {D.}~\bibnamefont {Sank}}, \bibinfo {author} {\bibfnamefont {M.}~\bibnamefont {Weides}}, \bibinfo {author} {\bibfnamefont {J.}~\bibnamefont {Wenner}}, \bibinfo {author} {\bibfnamefont {A.~N.}\ \bibnamefont {Cleland}},\ and\ \bibinfo {author} {\bibfnamefont {J.~M.}\ \bibnamefont {Martinis}},\ }\href {https://doi.org/10.1103/PhysRevLett.103.200404} {\bibfield  {journal} {\bibinfo  {journal} {Physical Review Letters}\ }\textbf {\bibinfo {volume} {103}},\ \bibinfo {pages} {200404} (\bibinfo {year} {2009})}\BibitemShut
  {NoStop}%
\bibitem [{\citenamefont {Zurek}(2001)}]{zurekSubPlanckStructurePhasespaceits2001}%
  \BibitemOpen
  \bibfield  {author} {\bibinfo {author} {\bibfnamefont {W.~H.}\ \bibnamefont {Zurek}},\ }\href {https://doi.org/10.1038/35089017} {\bibfield  {journal} {\bibinfo  {journal} {Nature}\ }\textbf {\bibinfo {volume} {412}},\ \bibinfo {pages} {712} (\bibinfo {year} {2001})}\BibitemShut {NoStop}%
\bibitem [{\citenamefont {Ghosh}\ \emph {et~al.}(2009)\citenamefont {Ghosh}, \citenamefont {Roy}, \citenamefont {Genes},\ and\ \citenamefont {Vitali}}]{ghoshSubPlanckscaleStructuresVibratingmoleculepresence2009}%
  \BibitemOpen
  \bibfield  {author} {\bibinfo {author} {\bibfnamefont {S.}~\bibnamefont {Ghosh}}, \bibinfo {author} {\bibfnamefont {U.}~\bibnamefont {Roy}}, \bibinfo {author} {\bibfnamefont {C.}~\bibnamefont {Genes}},\ and\ \bibinfo {author} {\bibfnamefont {D.}~\bibnamefont {Vitali}},\ }\href {https://doi.org/10.1103/PhysRevA.79.052104} {\bibfield  {journal} {\bibinfo  {journal} {Physical Review A}\ }\textbf {\bibinfo {volume} {79}},\ \bibinfo {pages} {052104} (\bibinfo {year} {2009})}\BibitemShut {NoStop}%
\bibitem [{\citenamefont {Kenfack}\ and\ \citenamefont {Yczkowski}(2004)}]{kenfackNegativityWignerFunctionindicatornonclassicality2004}%
  \BibitemOpen
  \bibfield  {author} {\bibinfo {author} {\bibfnamefont {A.}~\bibnamefont {Kenfack}}\ and\ \bibinfo {author} {\bibfnamefont {K.}~\bibnamefont {Yczkowski}},\ }\href {https://doi.org/10.1088/1464-4266/6/10/003} {\bibfield  {journal} {\bibinfo  {journal} {Journal of Optics B: Quantum and Semiclassical Optics}\ }\textbf {\bibinfo {volume} {6}},\ \bibinfo {pages} {396} (\bibinfo {year} {2004})}\BibitemShut {NoStop}%
\bibitem [{\citenamefont {Liao}\ \emph {et~al.}(2024)\citenamefont {Liao}, \citenamefont {Wang}, \citenamefont {Sitdikov}, \citenamefont {Salcedo}, \citenamefont {Seif},\ and\ \citenamefont {Minev}}]{liaoMachineLearningPracticalquantumerror2024}%
  \BibitemOpen
  \bibfield  {author} {\bibinfo {author} {\bibfnamefont {H.}~\bibnamefont {Liao}}, \bibinfo {author} {\bibfnamefont {D.~S.}\ \bibnamefont {Wang}}, \bibinfo {author} {\bibfnamefont {I.}~\bibnamefont {Sitdikov}}, \bibinfo {author} {\bibfnamefont {C.}~\bibnamefont {Salcedo}}, \bibinfo {author} {\bibfnamefont {A.}~\bibnamefont {Seif}},\ and\ \bibinfo {author} {\bibfnamefont {Z.~K.}\ \bibnamefont {Minev}},\ }\href {https://doi.org/10.1038/s42256-024-00927-2} {\bibfield  {journal} {\bibinfo  {journal} {Nature Machine Intelligence}\ }\textbf {\bibinfo {volume} {6}},\ \bibinfo {pages} {1478} (\bibinfo {year} {2024})}\BibitemShut {NoStop}%
\bibitem [{\citenamefont {Czarnik}\ \emph {et~al.}(2025)\citenamefont {Czarnik}, \citenamefont {McKerns}, \citenamefont {Sornborger},\ and\ \citenamefont {Cincio}}]{czarnikImprovingEfficiencyLearningbasederrormitigation2025}%
  \BibitemOpen
  \bibfield  {author} {\bibinfo {author} {\bibfnamefont {P.}~\bibnamefont {Czarnik}}, \bibinfo {author} {\bibfnamefont {M.}~\bibnamefont {McKerns}}, \bibinfo {author} {\bibfnamefont {A.~T.}\ \bibnamefont {Sornborger}},\ and\ \bibinfo {author} {\bibfnamefont {L.}~\bibnamefont {Cincio}},\ }\href {https://doi.org/10.22331/q-2025-05-05-1727} {\bibfield  {journal} {\bibinfo  {journal} {Quantum}\ }\textbf {\bibinfo {volume} {9}},\ \bibinfo {pages} {1727} (\bibinfo {year} {2025})}\BibitemShut {NoStop}%
\bibitem [{\citenamefont {Strikis}\ \emph {et~al.}(2021)\citenamefont {Strikis}, \citenamefont {Qin}, \citenamefont {Chen}, \citenamefont {Benjamin},\ and\ \citenamefont {Li}}]{strikisLearningBasedQuantumErrorMitigation2021}%
  \BibitemOpen
  \bibfield  {author} {\bibinfo {author} {\bibfnamefont {A.}~\bibnamefont {Strikis}}, \bibinfo {author} {\bibfnamefont {D.}~\bibnamefont {Qin}}, \bibinfo {author} {\bibfnamefont {Y.}~\bibnamefont {Chen}}, \bibinfo {author} {\bibfnamefont {S.~C.}\ \bibnamefont {Benjamin}},\ and\ \bibinfo {author} {\bibfnamefont {Y.}~\bibnamefont {Li}},\ }\href {https://doi.org/10.1103/PRXQuantum.2.040330} {\bibfield  {journal} {\bibinfo  {journal} {PRX Quantum}\ }\textbf {\bibinfo {volume} {2}},\ \bibinfo {pages} {040330} (\bibinfo {year} {2021})}\BibitemShut {NoStop}%
\bibitem [{\citenamefont {Liao}\ \emph {et~al.}(2025)\citenamefont {Liao}, \citenamefont {Zhu}, \citenamefont {Chiribella},\ and\ \citenamefont {Yang}}]{liaoNoiseagnosticQuantumErrormitigationdata2025}%
  \BibitemOpen
  \bibfield  {author} {\bibinfo {author} {\bibfnamefont {M.}~\bibnamefont {Liao}}, \bibinfo {author} {\bibfnamefont {Y.}~\bibnamefont {Zhu}}, \bibinfo {author} {\bibfnamefont {G.}~\bibnamefont {Chiribella}},\ and\ \bibinfo {author} {\bibfnamefont {Y.}~\bibnamefont {Yang}},\ }\href {https://doi.org/10.1038/s41534-025-00960-y} {\bibfield  {journal} {\bibinfo  {journal} {npj Quantum Information}\ }\textbf {\bibinfo {volume} {11}},\ \bibinfo {pages} {8} (\bibinfo {year} {2025})}\BibitemShut {NoStop}%
\bibitem [{\citenamefont {Steele}\ \emph {et~al.}(2025)\citenamefont {Steele}, \citenamefont {Reising},\ and\ \citenamefont {Li}}]{steeleRecoveryQuantumCorrelationsusingmachine2025}%
  \BibitemOpen
  \bibfield  {author} {\bibinfo {author} {\bibfnamefont {E.~W.}\ \bibnamefont {Steele}}, \bibinfo {author} {\bibfnamefont {D.~R.}\ \bibnamefont {Reising}},\ and\ \bibinfo {author} {\bibfnamefont {T.}~\bibnamefont {Li}},\ }\href {https://doi.org/10.1103/PhysRevApplied.23.034083} {\bibfield  {journal} {\bibinfo  {journal} {Physical Review Applied}\ }\textbf {\bibinfo {volume} {23}},\ \bibinfo {pages} {034083} (\bibinfo {year} {2025})}\BibitemShut {NoStop}%
\bibitem [{\citenamefont {Aguiar}\ \emph {et~al.}(2025)\citenamefont {Aguiar}, \citenamefont {Wold}, \citenamefont {Denisov},\ and\ \citenamefont {Ribeiro}}]{aguiarQuantumLiouvillianTomography2025}%
  \BibitemOpen
  \bibfield  {author} {\bibinfo {author} {\bibfnamefont {D.}~\bibnamefont {Aguiar}}, \bibinfo {author} {\bibfnamefont {K.}~\bibnamefont {Wold}}, \bibinfo {author} {\bibfnamefont {S.}~\bibnamefont {Denisov}},\ and\ \bibinfo {author} {\bibfnamefont {P.}~\bibnamefont {Ribeiro}},\ }\href {https://doi.org/10.48550/ARXIV.2504.10393} {\bibinfo {title} {Quantum {{Liouvillian Tomography}}}} (\bibinfo {year} {2025})\BibitemShut {NoStop}%
\bibitem [{\citenamefont {Oliva}\ and\ \citenamefont {Steuernagel}(2019)}]{olivaQuantumKerrOscillatorsevolutionphase2019}%
  \BibitemOpen
  \bibfield  {author} {\bibinfo {author} {\bibfnamefont {M.}~\bibnamefont {Oliva}}\ and\ \bibinfo {author} {\bibfnamefont {O.}~\bibnamefont {Steuernagel}},\ }\href {https://doi.org/10.1103/PhysRevA.99.032104} {\bibfield  {journal} {\bibinfo  {journal} {Physical Review A}\ }\textbf {\bibinfo {volume} {99}},\ \bibinfo {pages} {032104} (\bibinfo {year} {2019})}\BibitemShut {NoStop}%
\bibitem [{\citenamefont {Tiunov}\ \emph {et~al.}(2020)\citenamefont {Tiunov}, \citenamefont {Tiunova~(Vyborova)}, \citenamefont {Ulanov}, \citenamefont {Lvovsky},\ and\ \citenamefont {Fedorov}}]{tiunovExperimentalQuantumHomodynetomographymachine2020}%
  \BibitemOpen
  \bibfield  {author} {\bibinfo {author} {\bibfnamefont {E.~S.}\ \bibnamefont {Tiunov}}, \bibinfo {author} {\bibfnamefont {V.~V.}\ \bibnamefont {Tiunova~(Vyborova)}}, \bibinfo {author} {\bibfnamefont {A.~E.}\ \bibnamefont {Ulanov}}, \bibinfo {author} {\bibfnamefont {A.~I.}\ \bibnamefont {Lvovsky}},\ and\ \bibinfo {author} {\bibfnamefont {A.~K.}\ \bibnamefont {Fedorov}},\ }\href {https://doi.org/10.1364/OPTICA.389482} {\bibfield  {journal} {\bibinfo  {journal} {Optica}\ }\textbf {\bibinfo {volume} {7}},\ \bibinfo {pages} {448} (\bibinfo {year} {2020})}\BibitemShut {NoStop}%
\bibitem [{\citenamefont {Varona}\ \emph {et~al.}(2025)\citenamefont {Varona}, \citenamefont {M{\"u}ller},\ and\ \citenamefont {Bermudez}}]{varonaLindbladlikeQuantumTomographynonMarkovianquantum2025}%
  \BibitemOpen
  \bibfield  {author} {\bibinfo {author} {\bibfnamefont {S.}~\bibnamefont {Varona}}, \bibinfo {author} {\bibfnamefont {M.}~\bibnamefont {M{\"u}ller}},\ and\ \bibinfo {author} {\bibfnamefont {A.}~\bibnamefont {Bermudez}},\ }\href {https://doi.org/10.1038/s41534-025-01044-7} {\bibfield  {journal} {\bibinfo  {journal} {npj Quantum Information}\ }\textbf {\bibinfo {volume} {11}},\ \bibinfo {pages} {96} (\bibinfo {year} {2025})}\BibitemShut {NoStop}%
\bibitem [{\citenamefont {White}\ \emph {et~al.}(2022)\citenamefont {White}, \citenamefont {Pollock}, \citenamefont {Hollenberg}, \citenamefont {Modi},\ and\ \citenamefont {Hill}}]{whiteNonMarkovianQuantumProcessTomography2022}%
  \BibitemOpen
  \bibfield  {author} {\bibinfo {author} {\bibfnamefont {G.}~\bibnamefont {White}}, \bibinfo {author} {\bibfnamefont {F.}~\bibnamefont {Pollock}}, \bibinfo {author} {\bibfnamefont {L.}~\bibnamefont {Hollenberg}}, \bibinfo {author} {\bibfnamefont {K.}~\bibnamefont {Modi}},\ and\ \bibinfo {author} {\bibfnamefont {C.}~\bibnamefont {Hill}},\ }\href {https://doi.org/10.1103/PRXQuantum.3.020344} {\bibfield  {journal} {\bibinfo  {journal} {PRX Quantum}\ }\textbf {\bibinfo {volume} {3}},\ \bibinfo {pages} {020344} (\bibinfo {year} {2022})}\BibitemShut {NoStop}%
\bibitem [{\citenamefont {Knigge}\ \emph {et~al.}()\citenamefont {Knigge}, \citenamefont {Romero}, \citenamefont {Gu}, \citenamefont {Gavves}, \citenamefont {Bekkers}, \citenamefont {Tomczak}, \citenamefont {Hoogendoorn},\ and\ \citenamefont {Sonke}}]{kniggeModellingLongRange2023}%
  \BibitemOpen
  \bibfield  {author} {\bibinfo {author} {\bibfnamefont {D.~M.}\ \bibnamefont {Knigge}}, \bibinfo {author} {\bibfnamefont {D.~W.}\ \bibnamefont {Romero}}, \bibinfo {author} {\bibfnamefont {A.}~\bibnamefont {Gu}}, \bibinfo {author} {\bibfnamefont {E.}~\bibnamefont {Gavves}}, \bibinfo {author} {\bibfnamefont {E.~J.}\ \bibnamefont {Bekkers}}, \bibinfo {author} {\bibfnamefont {J.~M.}\ \bibnamefont {Tomczak}}, \bibinfo {author} {\bibfnamefont {M.}~\bibnamefont {Hoogendoorn}},\ and\ \bibinfo {author} {\bibfnamefont {J.-J.}\ \bibnamefont {Sonke}},\ }\href {https://doi.org/10.48550/ARXIV.2301.10540} {\bibinfo {title} {Modelling long range dependencies in \$n\$d: From task-specific to a general purpose cnn}}\BibitemShut {NoStop}%
\bibitem [{\citenamefont {Ng}\ \emph {et~al.}()\citenamefont {Ng}, \citenamefont {Lopez-Rodriguez}, \citenamefont {Balntas},\ and\ \citenamefont {Mikolajczyk}}]{ngReassessingLimitationsCNN2021}%
  \BibitemOpen
  \bibfield  {author} {\bibinfo {author} {\bibfnamefont {T.}~\bibnamefont {Ng}}, \bibinfo {author} {\bibfnamefont {A.}~\bibnamefont {Lopez-Rodriguez}}, \bibinfo {author} {\bibfnamefont {V.}~\bibnamefont {Balntas}},\ and\ \bibinfo {author} {\bibfnamefont {K.}~\bibnamefont {Mikolajczyk}},\ }\href {https://doi.org/10.48550/ARXIV.2108.07260} {\bibinfo {title} {Reassessing the limitations of cnn methods for camera pose regression}}\BibitemShut {NoStop}%
\bibitem [{\citenamefont {Ronneberger}\ \emph {et~al.}(2015)\citenamefont {Ronneberger}, \citenamefont {Fischer},\ and\ \citenamefont {Brox}}]{ronnebergerUNetConvolutionalNetworks2015}%
  \BibitemOpen
  \bibfield  {author} {\bibinfo {author} {\bibfnamefont {O.}~\bibnamefont {Ronneberger}}, \bibinfo {author} {\bibfnamefont {P.}~\bibnamefont {Fischer}},\ and\ \bibinfo {author} {\bibfnamefont {T.}~\bibnamefont {Brox}},\ }in\ \href {https://doi.org/10.1007/978-3-319-24574-4_28} {\emph {\bibinfo {booktitle} {Medical Image Computing and Computer-Assisted Intervention -- MICCAI 2015}}},\ Vol.\ \bibinfo {volume} {9351}\ (\bibinfo  {publisher} {Springer International Publishing},\ \bibinfo {address} {Cham},\ \bibinfo {year} {2015})\ pp.\ \bibinfo {pages} {234--241}\BibitemShut {NoStop}%
\bibitem [{\citenamefont {Herde}\ \emph {et~al.}(2024)\citenamefont {Herde}, \citenamefont {Raoni{\'c}}, \citenamefont {Rohner}, \citenamefont {K{\"a}ppeli}, \citenamefont {Molinaro}, \citenamefont {{de B{\'e}zenac}},\ and\ \citenamefont {Mishra}}]{herdePoseidonEfficientFoundationModelsPDEs2024}%
  \BibitemOpen
  \bibfield  {author} {\bibinfo {author} {\bibfnamefont {M.}~\bibnamefont {Herde}}, \bibinfo {author} {\bibfnamefont {B.}~\bibnamefont {Raoni{\'c}}}, \bibinfo {author} {\bibfnamefont {T.}~\bibnamefont {Rohner}}, \bibinfo {author} {\bibfnamefont {R.}~\bibnamefont {K{\"a}ppeli}}, \bibinfo {author} {\bibfnamefont {R.}~\bibnamefont {Molinaro}}, \bibinfo {author} {\bibfnamefont {E.}~\bibnamefont {{de B{\'e}zenac}}},\ and\ \bibinfo {author} {\bibfnamefont {S.}~\bibnamefont {Mishra}},\ }\href {https://doi.org/10.48550/ARXIV.2405.19101} {\bibinfo {title} {Poseidon: {{Efficient Foundation Models}} for {{PDEs}}}} (\bibinfo {year} {2024})\BibitemShut {NoStop}%
\bibitem [{\citenamefont {Peebles}\ and\ \citenamefont {Xie}(2022)}]{peeblesScalableDiffusionModelsTransformers2022}%
  \BibitemOpen
  \bibfield  {author} {\bibinfo {author} {\bibfnamefont {W.}~\bibnamefont {Peebles}}\ and\ \bibinfo {author} {\bibfnamefont {S.}~\bibnamefont {Xie}},\ }\href {https://doi.org/10.48550/ARXIV.2212.09748} {\bibinfo {title} {Scalable {{Diffusion Models}} with {{Transformers}}}} (\bibinfo {year} {2022})\BibitemShut {NoStop}%
\bibitem [{\citenamefont {Song}\ \emph {et~al.}(2020)\citenamefont {Song}, \citenamefont {{Sohl-Dickstein}}, \citenamefont {Kingma}, \citenamefont {Kumar}, \citenamefont {Ermon},\ and\ \citenamefont {Poole}}]{songScoreBasedGenerativeModelingStochasticDifferential2020}%
  \BibitemOpen
  \bibfield  {author} {\bibinfo {author} {\bibfnamefont {Y.}~\bibnamefont {Song}}, \bibinfo {author} {\bibfnamefont {J.}~\bibnamefont {{Sohl-Dickstein}}}, \bibinfo {author} {\bibfnamefont {D.~P.}\ \bibnamefont {Kingma}}, \bibinfo {author} {\bibfnamefont {A.}~\bibnamefont {Kumar}}, \bibinfo {author} {\bibfnamefont {S.}~\bibnamefont {Ermon}},\ and\ \bibinfo {author} {\bibfnamefont {B.}~\bibnamefont {Poole}},\ }\href {https://doi.org/10.48550/ARXIV.2011.13456} {\bibinfo {title} {Score-{{Based Generative Modeling}} through {{Stochastic Differential Equations}}}} (\bibinfo {year} {2020})\BibitemShut {NoStop}%
\bibitem [{\citenamefont {Song}\ \emph {et~al.}()\citenamefont {Song}, \citenamefont {Lu}, \citenamefont {Xu}, \citenamefont {Zhu}, \citenamefont {Buckeridge},\ and\ \citenamefont {Li}}]{songTimelyGPTExtrapolatableTransformer2023}%
  \BibitemOpen
  \bibfield  {author} {\bibinfo {author} {\bibfnamefont {Z.}~\bibnamefont {Song}}, \bibinfo {author} {\bibfnamefont {Q.}~\bibnamefont {Lu}}, \bibinfo {author} {\bibfnamefont {H.}~\bibnamefont {Xu}}, \bibinfo {author} {\bibfnamefont {H.}~\bibnamefont {Zhu}}, \bibinfo {author} {\bibfnamefont {D.~L.}\ \bibnamefont {Buckeridge}},\ and\ \bibinfo {author} {\bibfnamefont {Y.}~\bibnamefont {Li}},\ }\href {https://doi.org/10.48550/ARXIV.2312.00817} {\bibinfo {title} {Timelygpt: Extrapolatable transformer pre-training for long-term time-series forecasting in healthcare}}\BibitemShut {NoStop}%
\bibitem [{\citenamefont {Sun}\ \emph {et~al.}()\citenamefont {Sun}, \citenamefont {Dong}, \citenamefont {Patra}, \citenamefont {Ma}, \citenamefont {Huang}, \citenamefont {Benhaim}, \citenamefont {Chaudhary}, \citenamefont {Song},\ and\ \citenamefont {Wei}}]{sunLengthExtrapolatableTransformer2023}%
  \BibitemOpen
  \bibfield  {author} {\bibinfo {author} {\bibfnamefont {Y.}~\bibnamefont {Sun}}, \bibinfo {author} {\bibfnamefont {L.}~\bibnamefont {Dong}}, \bibinfo {author} {\bibfnamefont {B.}~\bibnamefont {Patra}}, \bibinfo {author} {\bibfnamefont {S.}~\bibnamefont {Ma}}, \bibinfo {author} {\bibfnamefont {S.}~\bibnamefont {Huang}}, \bibinfo {author} {\bibfnamefont {A.}~\bibnamefont {Benhaim}}, \bibinfo {author} {\bibfnamefont {V.}~\bibnamefont {Chaudhary}}, \bibinfo {author} {\bibfnamefont {X.}~\bibnamefont {Song}},\ and\ \bibinfo {author} {\bibfnamefont {F.}~\bibnamefont {Wei}},\ }in\ \href {https://doi.org/10.18653/v1/2023.acl-long.816} {\emph {\bibinfo {booktitle} {Proceedings of the 61st Annual Meeting of the Association for Computational Linguistics (Volume 1: Long Papers)}}}\ (\bibinfo  {publisher} {Association for Computational Linguistics})\ pp.\ \bibinfo {pages} {14590--14604}\BibitemShut {NoStop}%
\bibitem [{\citenamefont {Manning}\ \emph {et~al.}(2008)\citenamefont {Manning}, \citenamefont {Raghavan},\ and\ \citenamefont {Sch{\"u}tze}}]{manningIntroductionInformationRetrieval2008}%
  \BibitemOpen
  \bibfield  {author} {\bibinfo {author} {\bibfnamefont {C.~D.}\ \bibnamefont {Manning}}, \bibinfo {author} {\bibfnamefont {P.}~\bibnamefont {Raghavan}},\ and\ \bibinfo {author} {\bibfnamefont {H.}~\bibnamefont {Sch{\"u}tze}},\ }\href {https://doi.org/10.1017/CBO9780511809071} {\emph {\bibinfo {title} {Introduction to {{Information Retrieval}}}}},\ \bibinfo {edition} {1st}\ ed.\ (\bibinfo  {publisher} {Cambridge University Press},\ \bibinfo {year} {2008})\BibitemShut {NoStop}%
\bibitem [{\citenamefont {Sellier}\ and\ \citenamefont {Dimov}(2015)}]{sellierSensitivityStudyWignerMonteCarlo2015}%
  \BibitemOpen
  \bibfield  {author} {\bibinfo {author} {\bibfnamefont {J.}~\bibnamefont {Sellier}}\ and\ \bibinfo {author} {\bibfnamefont {I.}~\bibnamefont {Dimov}},\ }\href {https://doi.org/10.1016/j.cam.2014.09.010} {\bibfield  {journal} {\bibinfo  {journal} {Journal of Computational and Applied Mathematics}\ }\textbf {\bibinfo {volume} {277}},\ \bibinfo {pages} {87} (\bibinfo {year} {2015})}\BibitemShut {NoStop}%
\bibitem [{\citenamefont {Nazir}\ and\ \citenamefont {Schaller}()}]{nazirReactionCoordinateMapping2018}%
  \BibitemOpen
  \bibfield  {author} {\bibinfo {author} {\bibfnamefont {A.}~\bibnamefont {Nazir}}\ and\ \bibinfo {author} {\bibfnamefont {G.}~\bibnamefont {Schaller}},\ }in\ \href {https://doi.org/10.1007/978-3-319-99046-0_23} {\emph {\bibinfo {booktitle} {Thermodynamics in the Quantum Regime}}},\ Vol.\ \bibinfo {volume} {195},\ \bibinfo {editor} {edited by\ \bibinfo {editor} {\bibfnamefont {F.}~\bibnamefont {Binder}}, \bibinfo {editor} {\bibfnamefont {L.~A.}\ \bibnamefont {Correa}}, \bibinfo {editor} {\bibfnamefont {C.}~\bibnamefont {Gogolin}}, \bibinfo {editor} {\bibfnamefont {J.}~\bibnamefont {Anders}},\ and\ \bibinfo {editor} {\bibfnamefont {G.}~\bibnamefont {Adesso}}}\ (\bibinfo  {publisher} {Springer International Publishing})\ pp.\ \bibinfo {pages} {551--577}\BibitemShut {NoStop}%
\bibitem [{\citenamefont {De~Vega}\ and\ \citenamefont {Alonso}(2017)}]{devegaDynamicsNonMarkovianOpenquantumsystems2017}%
  \BibitemOpen
  \bibfield  {author} {\bibinfo {author} {\bibfnamefont {I.}~\bibnamefont {De~Vega}}\ and\ \bibinfo {author} {\bibfnamefont {D.}~\bibnamefont {Alonso}},\ }\href {https://doi.org/10.1103/RevModPhys.89.015001} {\bibfield  {journal} {\bibinfo  {journal} {Reviews of Modern Physics}\ }\textbf {\bibinfo {volume} {89}},\ \bibinfo {pages} {015001} (\bibinfo {year} {2017})}\BibitemShut {NoStop}%
\bibitem [{\citenamefont {Breuer}\ \emph {et~al.}(2009)\citenamefont {Breuer}, \citenamefont {Laine},\ and\ \citenamefont {Piilo}}]{breuerMeasureDegreeNonMarkovianBehaviorQuantum2009}%
  \BibitemOpen
  \bibfield  {author} {\bibinfo {author} {\bibfnamefont {H.-P.}\ \bibnamefont {Breuer}}, \bibinfo {author} {\bibfnamefont {E.-M.}\ \bibnamefont {Laine}},\ and\ \bibinfo {author} {\bibfnamefont {J.}~\bibnamefont {Piilo}},\ }\href {https://doi.org/10.1103/PhysRevLett.103.210401} {\bibfield  {journal} {\bibinfo  {journal} {Physical Review Letters}\ }\textbf {\bibinfo {volume} {103}},\ \bibinfo {pages} {210401} (\bibinfo {year} {2009})}\BibitemShut {NoStop}%
\bibitem [{\citenamefont {Rivas}\ \emph {et~al.}(2010)\citenamefont {Rivas}, \citenamefont {Huelga},\ and\ \citenamefont {Plenio}}]{rivasEntanglementNonMarkovianityQuantumEvolutions2010}%
  \BibitemOpen
  \bibfield  {author} {\bibinfo {author} {\bibfnamefont {{\'A}.}~\bibnamefont {Rivas}}, \bibinfo {author} {\bibfnamefont {S.~F.}\ \bibnamefont {Huelga}},\ and\ \bibinfo {author} {\bibfnamefont {M.~B.}\ \bibnamefont {Plenio}},\ }\href {https://doi.org/10.1103/PhysRevLett.105.050403} {\bibfield  {journal} {\bibinfo  {journal} {Physical Review Letters}\ }\textbf {\bibinfo {volume} {105}},\ \bibinfo {pages} {050403} (\bibinfo {year} {2010})}\BibitemShut {NoStop}%
\bibitem [{\citenamefont {Hu}\ \emph {et~al.}(1992)\citenamefont {Hu}, \citenamefont {Paz},\ and\ \citenamefont {Zhang}}]{huQuantumBrownianMotiongeneralenvironment1992}%
  \BibitemOpen
  \bibfield  {author} {\bibinfo {author} {\bibfnamefont {B.~L.}\ \bibnamefont {Hu}}, \bibinfo {author} {\bibfnamefont {J.~P.}\ \bibnamefont {Paz}},\ and\ \bibinfo {author} {\bibfnamefont {Y.}~\bibnamefont {Zhang}},\ }\href {https://doi.org/10.1103/PhysRevD.45.2843} {\bibfield  {journal} {\bibinfo  {journal} {Physical Review D}\ }\textbf {\bibinfo {volume} {45}},\ \bibinfo {pages} {2843} (\bibinfo {year} {1992})}\BibitemShut {NoStop}%
\bibitem [{\citenamefont {Richter}\ and\ \citenamefont {Breuer}(2024)}]{richterPhasespaceMeasuresInformationflowopen2024}%
  \BibitemOpen
  \bibfield  {author} {\bibinfo {author} {\bibfnamefont {M.~F.}\ \bibnamefont {Richter}}\ and\ \bibinfo {author} {\bibfnamefont {H.-P.}\ \bibnamefont {Breuer}},\ }\href {https://doi.org/10.1103/PhysRevA.110.062401} {\bibfield  {journal} {\bibinfo  {journal} {Physical Review A}\ }\textbf {\bibinfo {volume} {110}},\ \bibinfo {pages} {062401} (\bibinfo {year} {2024})}\BibitemShut {NoStop}%
\bibitem [{\citenamefont {{Ch{\'a}vez-Carlos}}\ \emph {et~al.}(2024)\citenamefont {{Ch{\'a}vez-Carlos}}, \citenamefont {{Garrido-Ram{\'i}rez}}, \citenamefont {Carmona}, \citenamefont {Batista}, \citenamefont {{Trallero-Herrero}}, \citenamefont {{P{\'e}rez-Bernal}}, \citenamefont {{Bastarrachea-Magnani}},\ and\ \citenamefont {Santos}}]{chavez-carlosQuantumSensingKerrparametricoscillators2024}%
  \BibitemOpen
  \bibfield  {author} {\bibinfo {author} {\bibfnamefont {J.}~\bibnamefont {{Ch{\'a}vez-Carlos}}}, \bibinfo {author} {\bibfnamefont {D.}~\bibnamefont {{Garrido-Ram{\'i}rez}}}, \bibinfo {author} {\bibfnamefont {A.~J.~V.}\ \bibnamefont {Carmona}}, \bibinfo {author} {\bibfnamefont {V.~S.}\ \bibnamefont {Batista}}, \bibinfo {author} {\bibfnamefont {C.~A.}\ \bibnamefont {{Trallero-Herrero}}}, \bibinfo {author} {\bibfnamefont {F.}~\bibnamefont {{P{\'e}rez-Bernal}}}, \bibinfo {author} {\bibfnamefont {M.~A.}\ \bibnamefont {{Bastarrachea-Magnani}}},\ and\ \bibinfo {author} {\bibfnamefont {L.~F.}\ \bibnamefont {Santos}},\ }\href {https://doi.org/10.48550/arXiv.2407.14590} {\bibinfo {title} {Quantum sensing in {{Kerr}} parametric oscillators}} (\bibinfo {year} {2024}),\ \Eprint {https://arxiv.org/abs/2407.14590} {arXiv:2407.14590 [quant-ph]} \BibitemShut {NoStop}%
\bibitem [{\citenamefont {Cai}\ \emph {et~al.}(2025)\citenamefont {Cai}, \citenamefont {Deng}, \citenamefont {Zhang}, \citenamefont {Ni}, \citenamefont {Mai}, \citenamefont {Huang}, \citenamefont {Zheng}, \citenamefont {Hu}, \citenamefont {Liu}, \citenamefont {Xu},\ and\ \citenamefont {Yu}}]{caiQuantumSqueezingAmplificationweakKerr2025}%
  \BibitemOpen
  \bibfield  {author} {\bibinfo {author} {\bibfnamefont {Y.}~\bibnamefont {Cai}}, \bibinfo {author} {\bibfnamefont {X.}~\bibnamefont {Deng}}, \bibinfo {author} {\bibfnamefont {L.}~\bibnamefont {Zhang}}, \bibinfo {author} {\bibfnamefont {Z.}~\bibnamefont {Ni}}, \bibinfo {author} {\bibfnamefont {J.}~\bibnamefont {Mai}}, \bibinfo {author} {\bibfnamefont {P.}~\bibnamefont {Huang}}, \bibinfo {author} {\bibfnamefont {P.}~\bibnamefont {Zheng}}, \bibinfo {author} {\bibfnamefont {L.}~\bibnamefont {Hu}}, \bibinfo {author} {\bibfnamefont {S.}~\bibnamefont {Liu}}, \bibinfo {author} {\bibfnamefont {Y.}~\bibnamefont {Xu}},\ and\ \bibinfo {author} {\bibfnamefont {D.}~\bibnamefont {Yu}},\ }\href {https://doi.org/10.1038/s41467-025-67699-0} {\bibfield  {journal} {\bibinfo  {journal} {Nature Communications}\ }\textbf {\bibinfo {volume} {17}},\ \bibinfo {pages} {970} (\bibinfo {year} {2025})}\BibitemShut {NoStop}%
\bibitem [{\citenamefont {Guo}\ \emph {et~al.}(2024)\citenamefont {Guo}, \citenamefont {He},\ and\ \citenamefont {Fadel}}]{guoQuantumMetrologySqueezedKerroscillator2024}%
  \BibitemOpen
  \bibfield  {author} {\bibinfo {author} {\bibfnamefont {J.}~\bibnamefont {Guo}}, \bibinfo {author} {\bibfnamefont {Q.}~\bibnamefont {He}},\ and\ \bibinfo {author} {\bibfnamefont {M.}~\bibnamefont {Fadel}},\ }\href {https://doi.org/10.1103/PhysRevA.109.052604} {\bibfield  {journal} {\bibinfo  {journal} {Physical Review A}\ }\textbf {\bibinfo {volume} {109}},\ \bibinfo {pages} {052604} (\bibinfo {year} {2024})}\BibitemShut {NoStop}%
\bibitem [{\citenamefont {Becker}\ \emph {et~al.}(2024)\citenamefont {Becker}, \citenamefont {Datta}, \citenamefont {Lami},\ and\ \citenamefont {Rouze}}]{beckerClassicalShadowTomographyContinuousVariables2024}%
  \BibitemOpen
  \bibfield  {author} {\bibinfo {author} {\bibfnamefont {S.}~\bibnamefont {Becker}}, \bibinfo {author} {\bibfnamefont {N.}~\bibnamefont {Datta}}, \bibinfo {author} {\bibfnamefont {L.}~\bibnamefont {Lami}},\ and\ \bibinfo {author} {\bibfnamefont {C.}~\bibnamefont {Rouze}},\ }\href {https://doi.org/10.1109/TIT.2024.3357972} {\bibfield  {journal} {\bibinfo  {journal} {IEEE Transactions on Information Theory}\ }\textbf {\bibinfo {volume} {70}},\ \bibinfo {pages} {3427} (\bibinfo {year} {2024})}\BibitemShut {NoStop}%
\bibitem [{\citenamefont {Wu}\ \emph {et~al.}(2024)\citenamefont {Wu}, \citenamefont {Zhu}, \citenamefont {Chiribella},\ and\ \citenamefont {Liu}}]{wuEfficientLearningContinuousvariablequantumstates2024}%
  \BibitemOpen
  \bibfield  {author} {\bibinfo {author} {\bibfnamefont {Y.-D.}\ \bibnamefont {Wu}}, \bibinfo {author} {\bibfnamefont {Y.}~\bibnamefont {Zhu}}, \bibinfo {author} {\bibfnamefont {G.}~\bibnamefont {Chiribella}},\ and\ \bibinfo {author} {\bibfnamefont {N.}~\bibnamefont {Liu}},\ }\href {https://doi.org/10.1103/PhysRevResearch.6.033280} {\bibfield  {journal} {\bibinfo  {journal} {Physical Review Research}\ }\textbf {\bibinfo {volume} {6}},\ \bibinfo {pages} {033280} (\bibinfo {year} {2024})}\BibitemShut {NoStop}%
\bibitem [{\citenamefont {Gandhari}\ \emph {et~al.}(2024)\citenamefont {Gandhari}, \citenamefont {Albert}, \citenamefont {Gerrits}, \citenamefont {Taylor},\ and\ \citenamefont {Gullans}}]{gandhariPrecisionBoundsContinuousVariableStateTomography2024}%
  \BibitemOpen
  \bibfield  {author} {\bibinfo {author} {\bibfnamefont {S.}~\bibnamefont {Gandhari}}, \bibinfo {author} {\bibfnamefont {V.~V.}\ \bibnamefont {Albert}}, \bibinfo {author} {\bibfnamefont {T.}~\bibnamefont {Gerrits}}, \bibinfo {author} {\bibfnamefont {J.~M.}\ \bibnamefont {Taylor}},\ and\ \bibinfo {author} {\bibfnamefont {M.~J.}\ \bibnamefont {Gullans}},\ }\href {https://doi.org/10.1103/PRXQuantum.5.010346} {\bibfield  {journal} {\bibinfo  {journal} {PRX Quantum}\ }\textbf {\bibinfo {volume} {5}},\ \bibinfo {pages} {010346} (\bibinfo {year} {2024})}\BibitemShut {NoStop}%
\bibitem [{\citenamefont {He}\ \emph {et~al.}(2024)\citenamefont {He}, \citenamefont {Yuan}, \citenamefont {Wong}, \citenamefont {Chakram}, \citenamefont {Seif}, \citenamefont {Jiang},\ and\ \citenamefont {Schuster}}]{heEfficientMultimodeWignertomography2024}%
  \BibitemOpen
  \bibfield  {author} {\bibinfo {author} {\bibfnamefont {K.}~\bibnamefont {He}}, \bibinfo {author} {\bibfnamefont {M.}~\bibnamefont {Yuan}}, \bibinfo {author} {\bibfnamefont {Y.}~\bibnamefont {Wong}}, \bibinfo {author} {\bibfnamefont {S.}~\bibnamefont {Chakram}}, \bibinfo {author} {\bibfnamefont {A.}~\bibnamefont {Seif}}, \bibinfo {author} {\bibfnamefont {L.}~\bibnamefont {Jiang}},\ and\ \bibinfo {author} {\bibfnamefont {D.~I.}\ \bibnamefont {Schuster}},\ }\href {https://doi.org/10.1038/s41467-024-48573-x} {\bibfield  {journal} {\bibinfo  {journal} {Nature Communications}\ }\textbf {\bibinfo {volume} {15}},\ \bibinfo {pages} {4138} (\bibinfo {year} {2024})}\BibitemShut {NoStop}%
\bibitem [{\citenamefont {Hradil}(1997)}]{hradilQuantumstateEstimation1997}%
  \BibitemOpen
  \bibfield  {author} {\bibinfo {author} {\bibfnamefont {Z.}~\bibnamefont {Hradil}},\ }\href {https://doi.org/10.1103/PhysRevA.55.R1561} {\bibfield  {journal} {\bibinfo  {journal} {Physical Review A}\ }\textbf {\bibinfo {volume} {55}},\ \bibinfo {pages} {R1561} (\bibinfo {year} {1997})}\BibitemShut {NoStop}%
\bibitem [{\citenamefont {{\v R}eh{\'a}{\v c}ek}\ \emph {et~al.}(2007)\citenamefont {{\v R}eh{\'a}{\v c}ek}, \citenamefont {Hradil}, \citenamefont {Knill},\ and\ \citenamefont {Lvovsky}}]{rehacekDilutedMaximumlikelihoodAlgorithmquantumtomography2007}%
  \BibitemOpen
  \bibfield  {author} {\bibinfo {author} {\bibfnamefont {J.}~\bibnamefont {{\v R}eh{\'a}{\v c}ek}}, \bibinfo {author} {\bibfnamefont {Z.}~\bibnamefont {Hradil}}, \bibinfo {author} {\bibfnamefont {E.}~\bibnamefont {Knill}},\ and\ \bibinfo {author} {\bibfnamefont {A.~I.}\ \bibnamefont {Lvovsky}},\ }\href {https://doi.org/10.1103/PhysRevA.75.042108} {\bibfield  {journal} {\bibinfo  {journal} {Physical Review A}\ }\textbf {\bibinfo {volume} {75}},\ \bibinfo {pages} {042108} (\bibinfo {year} {2007})}\BibitemShut {NoStop}%
\bibitem [{\citenamefont {Koutn{\'y}}\ \emph {et~al.}(2022)\citenamefont {Koutn{\'y}}, \citenamefont {Motka}, \citenamefont {Hradil}, \citenamefont {{\v R}eh{\'a}{\v c}ek},\ and\ \citenamefont {{S{\'a}nchez-Soto}}}]{koutnyNeuralnetworkQuantumStatetomography2022}%
  \BibitemOpen
  \bibfield  {author} {\bibinfo {author} {\bibfnamefont {D.}~\bibnamefont {Koutn{\'y}}}, \bibinfo {author} {\bibfnamefont {L.}~\bibnamefont {Motka}}, \bibinfo {author} {\bibfnamefont {Z.}~\bibnamefont {Hradil}}, \bibinfo {author} {\bibfnamefont {J.}~\bibnamefont {{\v R}eh{\'a}{\v c}ek}},\ and\ \bibinfo {author} {\bibfnamefont {L.~L.}\ \bibnamefont {{S{\'a}nchez-Soto}}},\ }\href {https://doi.org/10.1103/PhysRevA.106.012409} {\bibfield  {journal} {\bibinfo  {journal} {Physical Review A}\ }\textbf {\bibinfo {volume} {106}},\ \bibinfo {pages} {012409} (\bibinfo {year} {2022})}\BibitemShut {NoStop}%
\bibitem [{\citenamefont {Feng}\ \emph {et~al.}(2025)\citenamefont {Feng}, \citenamefont {Tao},\ and\ \citenamefont {Chen}}]{fengPhysicsInformedNeuralNetworksAdaptiveConstraints2025}%
  \BibitemOpen
  \bibfield  {author} {\bibinfo {author} {\bibfnamefont {C.}~\bibnamefont {Feng}}, \bibinfo {author} {\bibfnamefont {L.}~\bibnamefont {Tao}},\ and\ \bibinfo {author} {\bibfnamefont {L.}~\bibnamefont {Chen}},\ }\href {https://doi.org/10.48550/ARXIV.2512.14543} {\bibinfo {title} {Physics-{{Informed Neural Networks}} with {{Adaptive Constraints}} for {{Multi-Qubit Quantum Tomography}}}} (\bibinfo {year} {2025})\BibitemShut {NoStop}%
\end{thebibliography}

\clearpage
\foreach \n in {1,...,5} {
    \clearpage
    \thispagestyle{plain}
    \vspace*{-2.5cm} 
    \noindent\makebox[\textwidth][c]{%
        \includegraphics[width=0.99\paperwidth, page=\n]{figure/sm.pdf}%
    }
}
\end{document}